\documentclass[12pt]{iopart}
\usepackage[T1]{fontenc}
\usepackage[utf8]{inputenc}

\usepackage[dvipsnames]{xcolor}
\definecolor{linkcolor}{rgb}{0.0,0.3,0.5}
\usepackage[unicode,colorlinks=true,linkcolor=linkcolor,urlcolor=linkcolor,citecolor=linkcolor,pdfusetitle]{hyperref}
\usepackage{cite}
\usepackage[all]{hypcap}
\usepackage{graphicx}
\usepackage{xspace}
\usepackage{mathrsfs}
\usepackage{pifont}
\usepackage{lmodern}
\usepackage{enumitem}

\usepackage{iopams}
\usepackage{hyperref}
\usepackage{orcidlink}

\expandafter\let\csname equation*\endcsname\relax
\expandafter\let\csname endequation*\endcsname\relax

\usepackage{amsmath,amsfonts,amssymb}
\usepackage{amsthm}
\usepackage{etoolbox}
\usepackage{relsize, exscale}

\usepackage{longtable}

\usepackage{microtype}

\newcommand{\orcid}[1]{\href{https://orcid.org/#1}{\includegraphics[width=10pt]{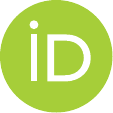}}}

\newcommand{\AEI}{Max Planck Institute for Gravitational Physics (Albert Einstein Institute), Am M\"uhlenberg 1, Potsdam 14476, Germany}
\newcommand{\Maryland}{Department of Physics, University of Maryland, College Park, MD 20742, USA}

\newcommand{\Cornell}{Cornell Center for Astrophysics and Planetary Science, Cornell University, Ithaca, New York 14853, USA}
\newcommand{\Caltech}{Theoretical Astrophysics 350-17, California Institute of Technology, Pasadena, California 91125, USA}
\newcommand{\Dartmouth}{Department of Mathematics, Center for Scientific Computing and Data Science Research, University of Massachusetts, Darmouth, MA 02747, USA}
\newcommand{\UIB}{Departament de F\'isica, Universitat de les Illes Balears, IAC3 -- IEEC, Crta. Valldemossa km 7.5, E-07122 Palma, Spain}

\begin{document}

\title{Impact of eccentricity and mean anomaly in numerical relativity mergers}

\author{%
Peter~James~Nee~\orcidlink{0000-0002-2362-5420}$^{1}$,
Aldo~Gamboa~\orcidlink{0000-0001-8391-5596}$^{1}$,
Harald~P.~Pfeiffer~\orcidlink{0000-0001-9288-519X}$^{1}$,
Lorenzo~Pompili~\orcidlink{0000-0002-0710-6778}$^{1}$,
Antoni~Ramos-Buades~\orcidlink{0000-0002-6874-7421}$^{2}$,
Vijay~Varma~\orcidlink{0000-0002-9994-1761}$^{3}$,
Michael~Boyle~\orcidlink{0000-0002-5075-5116}$^{4}$,
Alessandra~Buonanno~\orcidlink{0000-0002-5433-1409}$^{1,5}$,
Raffi~Enficiaud~\orcidlink{0000-0003-3908-1912}$^{1}$,
Lawrence~E.~Kidder~\orcidlink{0000-0001-5392-7342}$^{4}$
and
Mark~A.~Scheel~\orcidlink{0000-0001-6656-9134}$^{6}$
}

\address{$^{1}$~\AEI}
\address{$^{2}$~\UIB}
\address{$^{3}$~\Dartmouth}
\address{$^{4}$~\Cornell}
\address{$^{5}$~\Maryland}
\address{$^{6}$~\Caltech}

\ead{peter.nee@aei.mpg.de}

\hypersetup{pdfauthor={Peter James Nee et al.}}

\begin{abstract}
    Accurate modelling of black hole binaries is critical to
    achieve the science goals of gravitational-wave detectors.  Modelling
    such configurations relies strongly on calibration to numerical-relativity (NR)
    simulations.  Binaries on quasi-circular orbits have been
    widely explored in NR, however, coverage of the broader 9-dimensional parameter
    space, including orbital eccentricity, remains sparse.
    This article develops a new procedure
    to control orbital eccentricity of binary black hole simulations that
    enables choosing initial data parameters with precise control over
    eccentricity and mean anomaly of the subsequent evolution, as well as
    the coalescence time.  We then calculate several sequences of NR simulations that nearly uniformly cover the
    2-dimensional eccentricity--mean anomaly space for equal mass,
    non-spinning binary black holes.  We demonstrate that, for fixed
    eccentricity, many quantities related to the merger dynamics of
    binary black holes show an oscillatory dependence on mean anomaly.
    The amplitude of these oscillations scales
    nearly linearly with the eccentricity of the system. We find that for the
    eccentricities explored in this work, the magnitude of deviations in various quantities
    such as the merger amplitude and peak luminosity can approach $\sim5\%$
    of their quasi-circular value. We use our findings to explain
    eccentric phenomena reported in other studies.
    We also show that methods
    for estimating the remnant mass employed in the effective-one-body approach
    exhibit similar deviations, roughly matching the amplitude of the oscillations
    we find in NR simulations.
    This work is an important step towards a complete description of
    eccentric binary black hole mergers, and demonstrates the importance
    of considering the entire 2-dimensional parameter subspace related to eccentricity.
\end{abstract}

\maketitle

\section{Introduction}
By the end of the fourth observing run of the LIGO-Virgo-KAGRA network of gravitational-wave (GW) detectors, it is expected that the number of detected binary black hole (BBH) events will exceed 200~\cite{KAGRA:2013rdx, LIGOScientific:2018mvr, LIGOScientific:2019lzm,LIGOScientific:2020ibl, LIGOScientific:2021usb, LIGOScientific:2021djp}. In order to both detect potential events, as well as to ascertain their properties via parameter estimation \cite{LIGOScientific:2018jsj, Venumadhav:2019lyq, LIGOScientific:2020ibl,Nitz:2021zwj, Gupte:2024jfe}, it is essential that we have accurate waveform models across the entire BBH parameter space. This necessity only becomes more drastic when considering future, next-generation detector missions such as Einstein Telescope, LISA, and Cosmic Explorer \cite{Punturo:2010zz, Reitze:2019iox, LISA:2017pwj, Dhani:2024jja}, which will not only be more sensitive (and so more prone to bias due to waveform systematics), but will also cover new frequency bands, opening up the possibility to detect BBH events from new formation channels \cite{Antonini:2017ash, kozai1962secular, Zevin:2018kzq}.

While many waveform models leverage analytic prescriptions or perturbative approaches, almost all modern waveform models rely on calibration to numerical-relativity (NR) simulations (e.g.~\cite{Pompili:2023tna, Thompson:2023ase, Nagar:2018zoe}) or are directly constructed from NR simulations~\cite{Varma:2019csw}. This calibration can only be effective in regions of parameter space with sufficient simulation coverage. To date, much of the focus of NR simulations has been on quasi-circular systems, representing a 7 dimensional subspace of possible configurations. This is natural, as orbital eccentricity is radiated away during the inspiral of a binary~\cite{PhysRev.131.435}, and so by the time the system enters current detectors' frequency bands it is typically well-described by a quasi-circular system. Still, an understanding of eccentricity is vital for several science goals; to ensure that if there are eccentric signals we can properly identify them~\cite{LIGOScientific:2023lpe, Dhurkunde:2023qoe, Gadre:2024ndy, Clarke:2022fma}, to explore the formation scenarios for different binary systems~\cite{PortegiesZwart:1999nm, Mandel:2009nx,  Samsing:2013kua, Rodriguez:2018rmd, Fragione:2018vty, Zevin:2018kzq,Zevin:2021rtf}, as well as to disambiguate eccentric effects from potential deviations of General Relativity~\cite{Saini:2022igm, Narayan:2023vhm, Shaikh:2024wyn}. To this end, there are on-going efforts to build waveform models for eccentric systems~\cite{Islam:2021mha, Ramos-Buades:2023yhy, Gamboa:2024hli, Liu:2023ldr, Nagar:2024dzj, Gamba:2024cvy}.  However, most waveform models that do allow for orbital eccentricity in the inspiral still assume a quasi-circular merger model, with the exception being \texttt{NRSur2dq1Ecc} \cite{Islam:2021mha}. As such, both the calibration to NR and the assessment of these merger models requires extending our coverage to the eccentric parameter space.

\begin{figure}
    \centering
    \includegraphics[width=\linewidth]{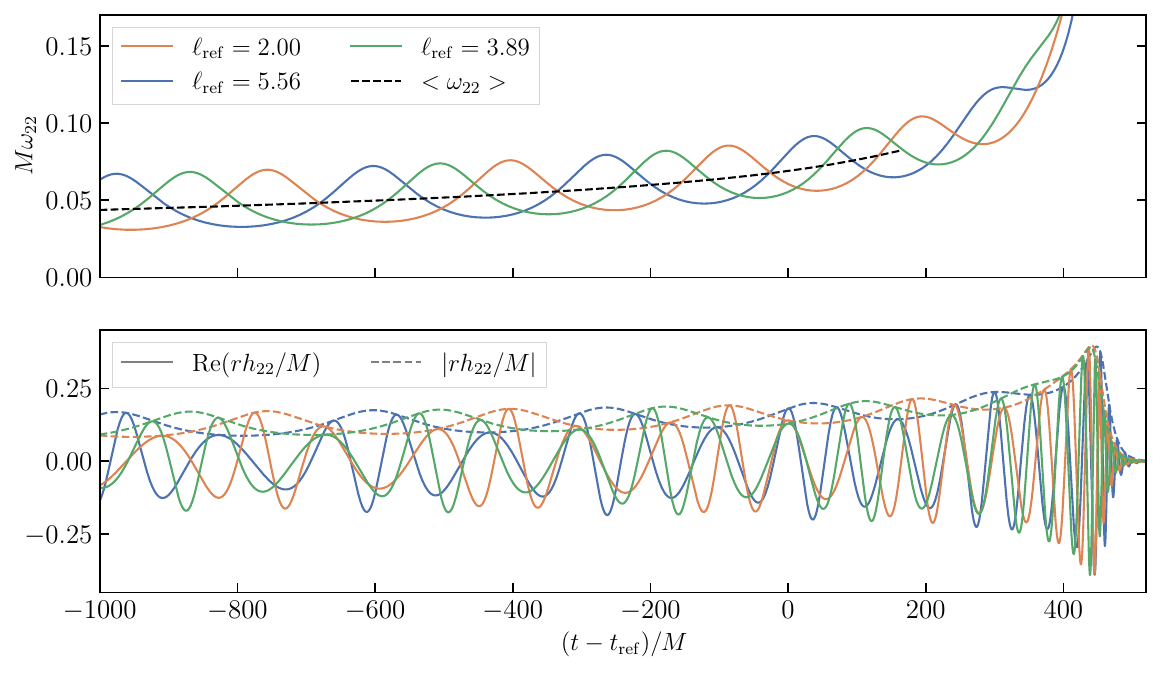}
    \caption{\textit{Importance of mean anomaly}.  \textbf{Top panel:} instantaneous GW frequency $\omega_{22}(t)$ for three NR simulations. The maxima and minima of these curves share common envelopes, indicating that these systems have the same eccentricity, and differ only in their mean anomaly. Note that all three systems share the same orbit-averaged frequency throughout their inspiral, plotted as the black dashed curve. \textbf{Bottom panel:} A representative GW mode, with phasing aligned at the time at which the orbit averaged frequency of the $(2,2)$ mode $\langle\omega_{22}\rangle=0.069/M$. While the systems accumulate the same overall number of GW cycles, the amplitude- and phase-modulations differ.  This plot zooms into the merger part of the NR simulations and the full simulations are visible in figure~\ref{plot: t_A22}.}
    \label{plot: t_omega22}
\end{figure}

At the same time, several NR groups have begun to expand into this part of the parameter space \cite{Hinder:2008kv, Mroue:2010re,Lewis:2016lgx, Hinder:2017sxy,Islam:2021mha, Huerta:2019oxn,Habib:2019cui, Ramos-Buades:2019uvh,Healy:2022wdn, Ramos-Buades:2022lgf, Wang:2023vka, Wang:2023wol, Wang:2024jro,Healy:2022wdn}. However, relaxing the quasi-circular restriction not only introduces a parameter describing the eccentricity of the system (namely, the eccentricity), but also a parameter describing its phasing (typically referred to as the anomaly, see figure~\ref{plot: t_omega22}). While these serve as excellent exploratory studies, to the authors' knowledge, no complete study accounting for both eccentric parameters has been performed thus far. As such, coverage of this 2-dimensional parameter space remains haphazard at best (see figure 3 of \cite{Islam:2021mha} for one such example).

While controlling the eccentricity of a simulation is a difficult problem in its own right \cite{Knapp:2024yww,Habib:2024soh}, controlling the anomaly at an arbitrary reference epoch proves even more challenging because for eccentric binary systems, the length of the simulation before the black holes (BHs) coalesce (referred to as the time to merger) has a much more sensitive dependence on the initial conditions compared to their quasi-circular counterparts. This sensitivity makes efficient exploration of eccentric binaries challenging. Therefore, to access such a complete study requires improvements upon our current eccentricity control methods.

In this work, we address two problems:  first, we propose a new eccentricity control method, which is based on the gravitational waveforms and utilizes state-of-the-art eccentric waveform models.  This new method not only allows exceptional control over the eccentricity of the binary, but also over the length of the simulations, which in turn enables targeting of specific values of the anomaly parameter. Second, we use this method to present the first complete exploration of the impact of both eccentricity and mean anomaly on several merger-related quantities, such as the peak of the strain and the remnant BH parameters. We demonstrate that many quantities have an oscillatory dependence on the mean anomaly, with the magnitude of these oscillations scaling roughly linearly with eccentricity. Using this information, we are able to explain features seen in previous eccentric studies. Finally, we show that some of these results may actually be accessible via analytic methods, using the \texttt{SEOBNRv5EHM} eccentric waveform model \cite{Gamboa:2024imd,Gamboa:2024hli}.\footnote{
    The \texttt{SEOBNRv5EHM} waveform model is publicly available through the Python package \texttt{pySEOBNR} (\url{https://waveforms.docs.ligo.org/software/pyseobnr}) \cite{Mihaylovv5}.
}

This paper is organised as follows: in Sec.~\ref{Sec:Methods}, we lay out several definitions used throughout this work, as well as our new waveform-based eccentricity control. In Sec.~\ref{results}, we demonstrate the effectiveness of this eccentricity control procedure, and use it to generate several sequences of eccentric simulations, which we use to assess the impact of eccentricity and mean anomaly. Finally, in Sec.~\ref{Sec:Conclusion}, we summarise our results, and discuss future avenues of research.

\section{Methods}
\label{Sec:Methods}

\subsection{Definitions}
\label{Sec:Definitions}

To avoid any gauge/definition ambiguities, we will define both eccentricity and mean anomaly directly from the gravitational waveform using the package \texttt{gw\_eccentricity} \cite{Shaikh:2023ypz, Ramos-Buades:2022lgf}. One first decomposes the complex waveform $h=h_+-ih_\times$ into a sum of spin-weighted spherical harmonics,
\begin{equation}
    h\left(t, \iota, \varphi_0\right)=\sum_{\ell=2}^{\ell=\infty} \sum_{m=-\ell}^{m=\ell} h_{\ell m}(t){ }_{-2} Y_{\ell m}\left(\iota, \varphi_0\right),
\end{equation}
where $\iota$ and $\varphi_0$ are the polar and azimuthal angles on the sky in the source frame, and ${ }_{-2} Y_{\ell m}$ are the $s=-2$ spin-weighted spherical harmonics. This study is largely restricted to binaries without orbital plane precession, and $\iota=0$ corresponds to the direction orthogonal to the orbital plane. Each mode $h_{\ell m}$ is further decomposed into a real amplitude and a real phase:

\begin{equation}
    h_{\ell m}\left(t\right) = A_{\ell m}\left(t\right)e^{-i\phi_{\ell m}\left(t\right)}.
\end{equation}

Following \cite{Shaikh:2023ypz, Ramos-Buades:2022lgf}, we introduce the frequency of $h_{22}$ as $\omega_{22} = d\phi_{22}/dt$, and define the eccentricity of the system as the envelope of $\omega_{22}$, i.e.,

\begin{equation}\label{eq: eccentricity}
    e_{\rm gw}(t)=\cos\left(\Psi/3\right)-\sqrt{3}\sin\left(\Psi/3\right),
\end{equation}
where
\begin{equation}
    \Psi=\arctan\left(\frac{1-e^2_{\omega_{22}}}{2e_{\omega_{22}}}\right), \, e_{\omega_{22}}(t)=\frac{\sqrt{\omega_{22}^{\mathrm{p}}(t)}-\sqrt{\omega_{22}^{\mathrm{a}}(t)}}{\sqrt{\omega_{22}^{\mathrm{p}}(t)}+\sqrt{\omega_{22}^{\mathrm{a}}(t)}},
\end{equation}
and $\omega_{22}^{\mathrm{a/p}}(t)$ are smooth interpolants of the $(2,2)$ mode frequency at apastron/periastron.  For a time interval between successive periastron passages $t_i^p<t<t_{i+1}^p$, i.e., between maxima of $\omega_{22}$, the mean anomaly is defined as
\begin{equation}
    \ell_{\mathrm{gw}}(t)=2 \pi \frac{t-t_i^{\mathrm{p}}}{t_{i+1}^{\mathrm{p}}-t_i^{\mathrm{p}}},
\end{equation}
where $t_i^{\mathrm{p}}$ is the time of the i'th periastron passage. Note that $\ell_{\mathrm{gw}}(t)$ is essentially a piecewise linear function in time, which increases from $0$ to $2\pi$ between each periastron passage. We will omit the subscript ``gw'' for the rest of this work.

As should be clear from the above formulae, both the eccentricity and mean anomaly of a given system are functions of time. As such, it is important to specify \textit{where} in the waveform we are defining eccentricity and mean anomaly when comparing systems. This is typically either done at some reference time $t_{\rm ref}$, or at some reference orbit-averaged frequency $\langle\omega_{22}\rangle(t)$, which we define as a smooth interpolant of

\begin{align}
    \left\langle\omega_{22}\right\rangle_i  =\frac{1}{t_{i+1}^{\mathrm{p}}-t_i^{\mathrm{p}}} \int_{t_i^{\mathrm{p}}}^{t_{i+1}^{\mathrm{p}}} \omega_{22}(t)\, \mathrm{d} t
    =\frac{\phi_{22}\left(t_{i+1}^{\mathrm{p}}\right)-\phi_{22}\left(t_i^{\mathrm{p}}\right)}{t_{i+1}^{\mathrm{p}}-t_i^{\mathrm{p}}}.
\end{align}

Throughout this work we will make use of both reference time and reference frequencies: for comparing systems of the same eccentricity we use reference frequencies, while comparing across eccentricities we will use reference times.
We will use subscripts to indicate when a quantity is defined (e.g., $\ell_{-700M}$ represents the mean anomaly $700M$ before merger, while $\ell_{0.03/M}$ would be the mean anomaly when $\langle\omega_{22}\rangle=0.03/M$). Additionally, we make frequent use of an orbit averaged frequency of $\langle\omega_{22}\rangle=0.069/M$, corresponding to an epoch near merger. For brevity, we will use the subscript ``ref'' (e.g. $\ell_{\rm ref}$ and $t_{\rm ref}$) to denote quantities evaluated at this epoch.

As will be discussed in Sec.~\ref{results}, the choice of exactly where in the inspiral to define the reference epoch is important when studying the impact of mean anomaly on merger dynamics.
The reason is that mean anomaly cycles through many periods during the inspiral (once per radial period), and so even a small secular dephasing between two runs can lead to dramatically different mean anomaly values.  In contrast, eccentricity is slowly and monotonically decaying, and a change in reference epoch for the definition of eccentricity will primarily result in a re-scaling of all eccentricities that respects their ordering (for constant BH parameters). However, when comparing across systems of varying BH parameters, one could have that the $e_i(t)$ curves cross, and so the choice of reference epoch becomes more important. As we will focus on equal mass non-spinning systems in this work, we leave careful consideration of the reference epoch for defining eccentricity to future work.

Figure~\ref{plot: t_omega22} gives a more intuitive picture of how eccentricity and mean anomaly paramaterize a system. We plot the evolution of $\omega_{22}$ for three sample NR simulations (all with mass-ratio $q=1$, both BHs non-spinning). As~Eq.~\eqref{eq: eccentricity} depends only on the apastron and periastron frequencies (i.e., the envelopes of the frequency), it should be clear that these three simulations share the same $e(t)$. As such, a second parameter is required to distinguish between these simulations, the ``phase'' of the radial orbit. This is precisely the mean anomaly $\ell(t)$.

All numerical simulations in this work were carried out using the Spectral Einstein Code (\texttt{SpEC}) \cite{SpECwebsite} developed by the Simulating eXtreme Spacetimes (SXS) collaboration. \texttt{SpEC} employs a multi-domain spectral discretization \cite{Kidder:1999fv, Scheel:2008rj, Szilagyi:2009qz, Hemberger:2012jz} to solve a first-order representation of the generalized harmonic system \cite{Lindblom:2005qh}. Excision surfaces are placed within apparent horizons \cite{Scheel:2008rj, Szilagyi:2009qz, Hemberger:2012jz, Ossokine:2013zga}, and constraint-preserving boundary conditions are used for the outer boundaries \cite{Lindblom:2005qh, Rinne:2006vv, Rinne:2007ui}. Initial data is constructed using \texttt{Spells} \cite{Pfeiffer:2002wt,Ossokine:2015yla}, which solves the extended conformal-thin sandwich (XCTS) equations \cite{York:1998hy,Pfeiffer:2002iy,Cook:2004kt}.

\subsection{Waveform-based eccentricity control}
\label{Sec:EOB-EccControl}

The initial data of a NR simulation determines the properties of
the system.  In \texttt{SpEC}, initial data is specified through
mass-ratio $q=m_1/m_2\ge 1$ and BH spins $\vec\chi_{1,2}$,
where the subscript labels the two BHs.  In this paper, we
will largely restrict our focus to BH spins parallel to the orbital angular momentum (which
in turn is parallel to the $\hat z$-axis), so that $\vec\chi_1=\hat
    z\chi_i$ where $\chi_i\in [-1,1]$ represents the projection of
$\vec\chi_i$ onto the orbital angular momentum vector.  In addition
to mass-ratio $q$ and spins $\chi_i$, one must also specify:
\begin{enumerate}
    \item the initial coordinate separation $D_0$,
    \item the initial orbital frequency $\Omega_0$, and
    \item the initial expansion $\dot{a}_0$, defined as the initial in-going velocity divided by the initial separation $v_{r}/D_0$.
\end{enumerate}
These three values encode the orbital properties (eccentricity \& radial phase) as well as the merger time $T_{\rm merger}$, which we define as the time of the peak of $A_{22}$.
Our goal is to obtain precise control over mean anomaly, eccentricity, and merger time. As such, we need a procedure to obtain initial parameters $D_0$, $\Omega_0$ and $\dot{a}_0$ that correspond to a NR simulation with the desired initial eccentric parameters ($e_0$, $\ell_0$), and time to merger $T_{\rm merger}$.

As a first step we will restrict ourselves to construct
initial data only at apastron. Because of the larger separation and smaller velocities of the BHs at apastron, junk radiation~\cite{Lovelace:2008hd} is reduced; moreover, this choice meshes well with technical restrictions of \texttt{SpEC} which limit by how much the separation can increase during an evolution~\cite{Boyle:2007ft}. We note that starting at apastron does not limit our ability to reach any physically possible configuration.  For instance, we achieve all three simulations shown in figure~\ref{plot: t_omega22} with initial data at apastron by suitably changing $D_0$, $\Omega_0$, and $\dot{a}_0$ so that the subsequent simulations have the same eccentricity (at a reference epoch) but slightly different durations between initial data and merger.  In Sec.~\ref{sec:31}, we will use this approach to systematically explore the mean anomaly parameter space.

We furthermore aim for an iterative procedure to adjust our initial conditions:  Based on evolutions lasting a few orbits, we aim to adjust our initial conditions, such that a subsequent evolution is closer to the desired configuration.  Many such iterative eccentricity control procedures have been employed to date~\cite{Pfeiffer:2007yz,Buchman:2012dw,Buonanno:2010yk,Purrer:2012wy,Ramos-Buades:2018azo,Habib:2020dba,Knapp:2024yww,Habib:2024soh},
however we shall improve on those in two ways:  First, we will use the gravitational waveform in our adjustment procedure, in order to eliminate the coordinate-dependence inherent in procedures that use the coordinate trajectories of the BHs~\cite{Pfeiffer:2007yz,Buonanno:2010yk,Ramos-Buades:2018azo,Habib:2020dba,Knapp:2024yww,Habib:2024soh}.
Second, in our updating step, we will utilize
state-of-the-art waveform models for eccentric BBH, rather than using a fitting function based on ad hoc choices (e.g.~\cite{Pfeiffer:2007yz,Buonanno:2010yk}) or post-Newtonian inspired approximations~\cite{Knapp:2024yww}.
Specifically, we employ \texttt{SEOBNRv5EHM}~\cite{Gamboa:2024hli,Gamboa:2024imd}, an aligned-spin,
effective-one-body (EOB) model for eccentric BBH systems. \texttt{SEOBNRv5EHM} parameterizes its orbit differently to NR initial data, namely with
\begin{enumerate}
    \item initial eccentricity $e_0$,
    \item initial relativistic anomaly $\zeta_0$,\footnote{
              The relativistic anomaly $ \zeta $ employed in the \texttt{SEOBNRv5EHM} model represents a different radial phase parameter than the mean anomaly $ \ell $. While the evolution of these parameters varies across radial orbits, they are defined such that both satisfy $ \zeta = \ell = 0 $ at periastron and $ \zeta = \ell = \pi $ at apastron.
          }
          and
    \item initial orbit-averaged orbital frequency $\langle\Omega\rangle_0$.
\end{enumerate}
The EOB model internally translates these quantities through a root-finding procedure into the actual initial conditions for the EOB dynamics evolution: separation $r_0$, instantaneous orbital frequency $\omega_0$, and radial momentum $p_r$.

\begin{figure}
    \centering
    \includegraphics[width=0.73\linewidth,trim=0 7 0 5]{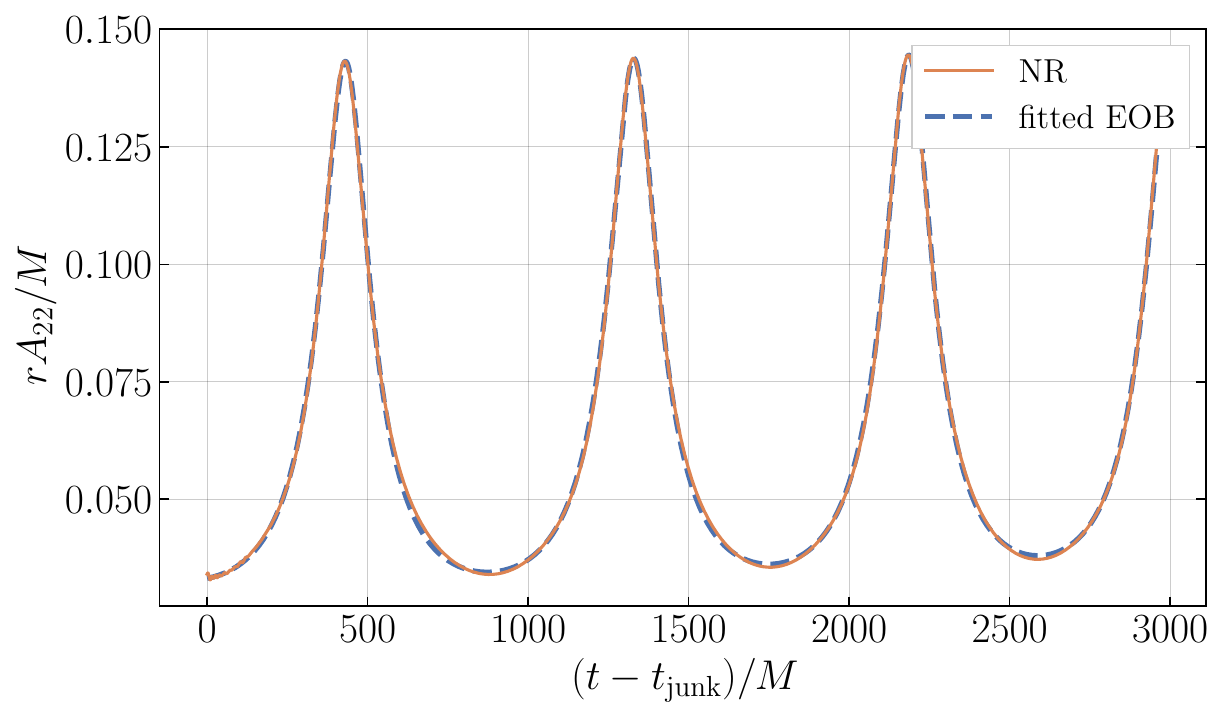}
    \caption{\textit{Waveform-based eccentricity control.} Shown is a single step of eccentricity control, where the blue curve corresponds to $A_{22}$ obtained via finite-radius extraction in an NR simulation, and the dashed orange curve corresponds to the best-fit \texttt{SEOBNRv5EHM} waveform. $t_{\rm junk}$ corresponds to the time at which junk has decreased by an acceptable amount for us to start fitting, which typically takes $\sim 100M$. System corresponds to a particular eccentricity control iteration for the tenth entry in table~\ref{table: ecc sequence}.
    }
    \label{plot: ecc control}
\end{figure}

Our task is now to combine \texttt{SEOBNRv5EHM} with short NR simulations to achieve our target configuration with initial eccentricity $e_0^{\rm target}$ and time to merger $T_{\rm merger}^{\rm target}$.  We seek an iterative procedure that ultimately yields  NR initial-data parameters $(D_0, \Omega_0, \dot{a}_0)$ for our target configuration.
Our eccentricity control method begins by constructing a first NR initial data as a seed for the subsequent iterative procedure:
\begin{enumerate}
    \item Fix $\zeta_0=0$, and perform a root find in the EOB initial parameters $(e_0, \langle\Omega\rangle_0)$ to find the EOB waveform with the desired eccentricity and waveform length.
    \item\label{step:EOB-target-params} Extract the initial dynamical quantities from the EOB model ($r_0,\omega_0,p_r$).
    \item Set the NR initial-data parameters to $D_0=r_0-M$, $\Omega_0=\omega_0$ and $\dot{a}_0=p_r/r_0$.
\end{enumerate}
Now that we have seed NR initial conditions, we can begin the actual iterative procedure:
\begin{enumerate}[resume]
    \item \label{step:EvolveNR}
          Construct the NR initial-data set for $(D_0, \Omega_0, \dot{a}_0)$, and evolve for three radial periods up to a time $t_{\rm max}$.
    \item Extract the gravitational waveform from the NR simulation.  We use a waveform extracted at a finite radius (typically $r\approx 300M$), discard data for $t<t_{\rm junk}$ which is contaminated by junk radiation, and calculate the amplitude of the $(2,2)$ mode, $A_{22}^{\rm NR}(t)$.
          Figure~\ref{plot: ecc control} plots an exemplary $A_{22}^{\rm NR}(t)$ obtained in this way; because we remove the initial junk-radiation phase, the data starts somewhat after an apastron passage, i.e., somewhat after a minimum in $A_{22}^{\rm NR}(t)$.
    \item \label{step:EOB-fit}Find the \texttt{SEOBNRv5EHM} waveform that best matches $A_{22}^{\rm NR}$.  In this step, we find EOB parameters $\left(e_0, \langle\Omega\rangle_0\right)$, a time-shift $\delta t$, and an overall amplitude offset $C$ to minimize
          \begin{equation}\label{eq:ObjectiveFunction}
              \int^{t_{\rm max}}_{t_{\rm junk}} \left| A_{22}^{\rm NR}(t)-A_{22}^{\rm EOB}(t+\delta t;e_{0}, \ell_\pm, \langle\Omega\rangle_0)+C \right| dt.
          \end{equation}
          The time-shift $\delta t$ aligns the time-axes of the EOB evolution with the NR evolution (e.g., to account for finite-radius GW extraction effects) and the overall offset $C$ is likewise necessary to reach good agreement and a robust fit in the light of using a finite-radius NR waveform.
          In addition,  for systems with low eccentricities, the EOB relativistic anomaly is varied over the two choices $\ell_\pm = \{0, \pi\}$ to account for rare cases that the NR simulation switches into a periastron-configuration, rather than the intended apastron-configuration (a more detailed explanation is provided below).
          Figure~\ref{plot: ecc control} also plots this best-fit EOB waveform.
    \item From the best-fit EOB waveform, obtain a predicted time to merger $T_{\rm merger}^{\rm fit}$ and initial eccentricity $e_0^{\rm fit}$.
          If $e_0^{\rm fit}$ and $T_{\rm merger}^{\rm fit}$ are sufficiently close to our target eccentricity $e_0^{\rm target}$ and time to merger $T_{\rm merger}^{\rm target}$ (within $\pm10^{-4}$ and $\pm{50M}$ respectively), and $\ell_\pm=\pi$, then the eccentricity control procedure was successful.  Exit the iterative loop and continue the NR simulation.
    \item Extract the initial EOB dynamical parameters  $\left(r_0^{\rm fit}, \omega_0^{\rm fit}, p_r^{\rm fit}\right)$ from the EOB fit obtained in step (\ref{step:EOB-fit}). Update the NR initial data parameters based on differences between these current EOB parameters and the target EOB parameters determined in step~(\ref{step:EOB-target-params}) above:
          \begin{eqnarray}
              D_0 &\leftarrow D_0 \frac{r_0}{r_0^{\rm fit}},\\
              \Omega_0 &\leftarrow \Omega_0 \frac{\omega_0}{\omega_0^{\rm fit}},\\
              \dot a_{0}&\leftarrow \dot a_0 \frac{p_r}{p_r^{\rm fit}}\frac{r_0^{\rm fit}}{r_0}.
          \end{eqnarray}
    \item    Go to step~(\ref{step:EvolveNR}).
\end{enumerate}

For large eccentricities, the inclusion of the fitted parameter $\ell_\pm$ is not required, as the behaviour of the binary at periastron and at apastron is quite distinct. However, for eccentricities $e\lesssim 10^{-2}$, the motion of the BHs is dominated by the monotonic inspiral as opposed to the eccentricity-driven oscillations. As such, it can happen that between eccentricity iterations, the NR simulation switches from starting at apastron to starting at periastron.  The same is true for the constant offset $C$: when the waveform is dominated by large, eccentricity-driven oscillations (e.g., as in figure~\ref{plot: ecc control}), the inclusion of $C$ in the fit makes little difference. However, for lower eccentricities, the near-zone effects become more noticeable and may degrade the quality of the fit.

\begin{table}
    \centering
    \label{table: ecc control cases}
    \begin{tabular}{l|c|c|c|c|c|c}
               & $q$ & $\vec\chi_1$  & $\vec\chi_2$ & $e_0^{\rm target}$ & $T_{\rm merger}^{\rm target}$ & $T_{\rm merger}$ \\ \hline
        Case 1 & 1   & $(0,0,0)$     & $(0,0,0)$    & $0.123$            & $11543M$                      & $11505M$         \\ \hline
        Case 2 & 2   & $(0,0,0.8)$   & $(0,0,0.8)$  & $0.496$            & $11688M$                      & $11468M$         \\ \hline
        Case 3 & 1   & $(0.7,0,0.4)$ & $(0.8,0,0)$  & $0.395$            & $14522M$                      & $13935M$         \\
    \end{tabular}
    \caption{Test cases for waveform-based eccentricity control procedure. The last column lists the actual time to merger obtained by the target NR simulation.}
\end{table}

Let us now investigate the efficacy of the iterative procedure just defined. We pick three representative cases, defined in Table~\ref{table: ecc control cases}. Specifically, we include a non-spinning system, an aligned spin system, as well as a precessing configuration.

\begin{figure}
    \centering
    \includegraphics[width=0.72\linewidth]{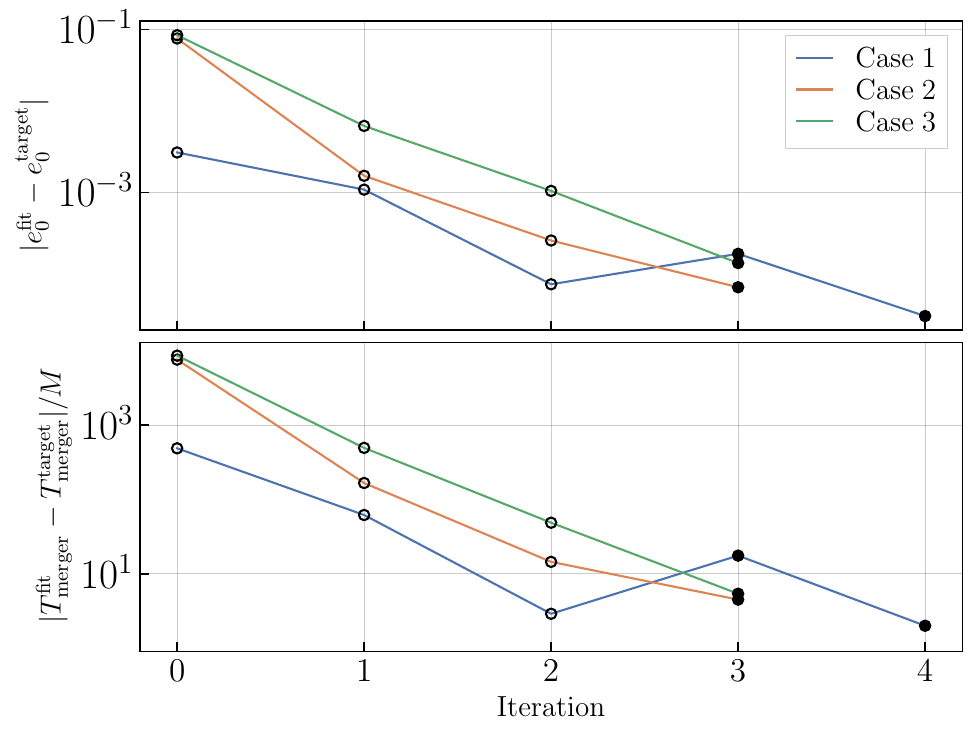}
    \caption{\textit{Performance of our eccentricity control procedure.} Plotted is the convergence of $(e^{\rm fit}_0, T^{\rm fit}_{\rm merger})$ for test cases defined in Table~\ref{table: ecc control cases}.  Open circles indicate iterations performed at low resolution to save computational time, while solid circles indicate iterations performed at production-level resolutions.  Note that $T^{\rm fit}_{\rm merger}$ is the estimated time to merger predicted by the best fit \texttt{SEOBNRv5EHM} system, not the actual time to merger.
    }
    \label{plot: ecc convergence}
\end{figure}

Figure~\ref{plot: ecc convergence} shows the convergence of
this iterative procedure for our test-configurations.
For improved computational efficiency,  we perform the first few iterations of eccentricity control at a numerical resolution lower than that we intend to use for our final simulation (referred to as ``rough eccentricity control'', and indicated by open circles in figure~\ref{plot: ecc convergence}).  The final iterations (filled circles) are run at production
resolutions. NR simulations at different numerical resolution will, in general, result in slightly different eccentricity. As
such, the switch from low to high resolution can move us away from our
target $(e_0^{\rm target}, T_{\rm merger}^{\rm target})$, as can be seen in the blue curve. We find that the
number of iterations required seems largely independent of the
parameters of the BBH, usually taking 3 or 4 iterations.

We emphasize that the convergence plotted is that of the \textit{expected} time to merger $T^{\rm fit}_{\rm merger}$, calculated from the EOB model.  This will, in general, differ from the merger time of the NR evolution by a few $100M$; see the last column in Table~\ref{table: ecc control cases}. There are several possible reasons for the discrepancy between $T_{\rm merger}$ and $T_{\rm merger}^{\rm fit}$; while it is possible this arises due to inaccuracies of the EOB model in parts of the parameter space, the results of Gamboa et al.~\cite{Gamboa:2024hli} indicate this is unlikely to be the dominant source of these errors. More likely is the fact that during step~(\ref{step:EOB-fit}), we are comparing an un-extrapolated, finite radius NR waveform with an EOB waveform associated with future null infinity. In any case, our procedure satisfies two important requirements:  First, the achieved $T_{\rm merger}$ is close enough to $T^{\rm target}_{\rm merger}$ for practical applications: Neither do the resulting NR simulations waste large amounts of CPU time simulating unneeded earlier parts of the inspiral, nor are the simulations accidentally much shorter than desired.   Secondly, $T^{\rm target}_{\rm merger}$ and $T_{\rm merger}$ correlate very well:  If we change $T^{\rm target}_{\rm merger}$ by a small fraction of an orbital period (say, $20M$), then the actual $T_{\rm merger}$ changes by a closely matching amount.  This property is essential in constructing sequences of simulations with nearly uniform spacing in mean anomaly.

In addition to the runs shown in figure~\ref{plot: ecc convergence}, we have also
tested the procedure further for spin-aligned systems with mass-ratios up to $q=6$, BH spins up to $\chi_i=0.8$, and eccentricities $10^{-3}\leq e_0\leq 0.5$.
For such eccentricities, we can achieve control over the target eccentricity $e_{0}^{\rm target}$ to better than $10^{-4}$. To extend this method to lower eccentricities (as well as to obtain more accurate control over the time to merger) will likely require switching from using finite-radius waveforms, to using extrapolated waveforms (either via partial extrapolation, or Cauchy Characteristic Extraction).

Finally, we have performed preliminary tests for precessing systems, an example of which is the green curve in figure~\ref{plot: ecc convergence}. While \texttt{SEOBNRv5EHM} contains no precession effects, we expect that $1)$ early in inspiral, for a short amount of orbits the precession effects are negligible, and $2)$ the time to merger from a given orbit is dominated by eccentricity dependence as opposed to non-aligned spin component dependence.
Since there is currently not an agreed-upon way of defining eccentricity from the gravitational waveform for precessing systems, it is difficult to provide an estimate of the precision of our eccentricity control. We find that for precessing systems we can control $e_0^{\rm fit}$ to be within $\pm 10^{-3}$ of $e_0^{\rm target}$ for eccentricities $\sim0.5$, while for a target $T^{\rm target}_{\rm merger}$ of approximately $12000M$ we can control the time to merger to within $\pm 500M$.

\section{Results}
\label{results}

\subsection{Two-dimensional parameter survey in $(e,\ell)$}
\label{sec:31}

\begin{figure*}
    \includegraphics[width=0.5\linewidth,trim=8 5 1 5, clip=true]{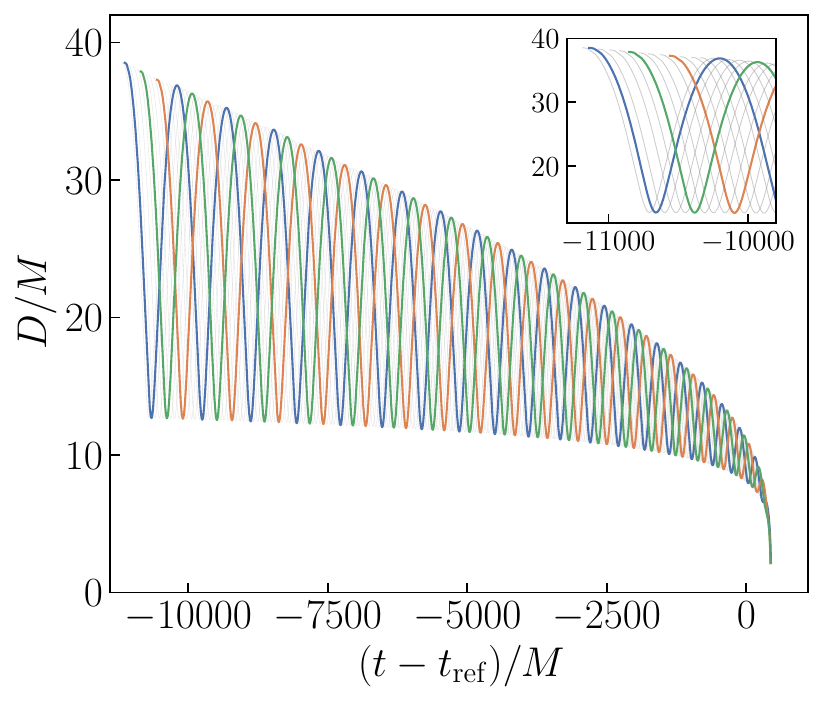}
    \includegraphics[width=0.5\linewidth,trim=1 5 4 5, clip=true]{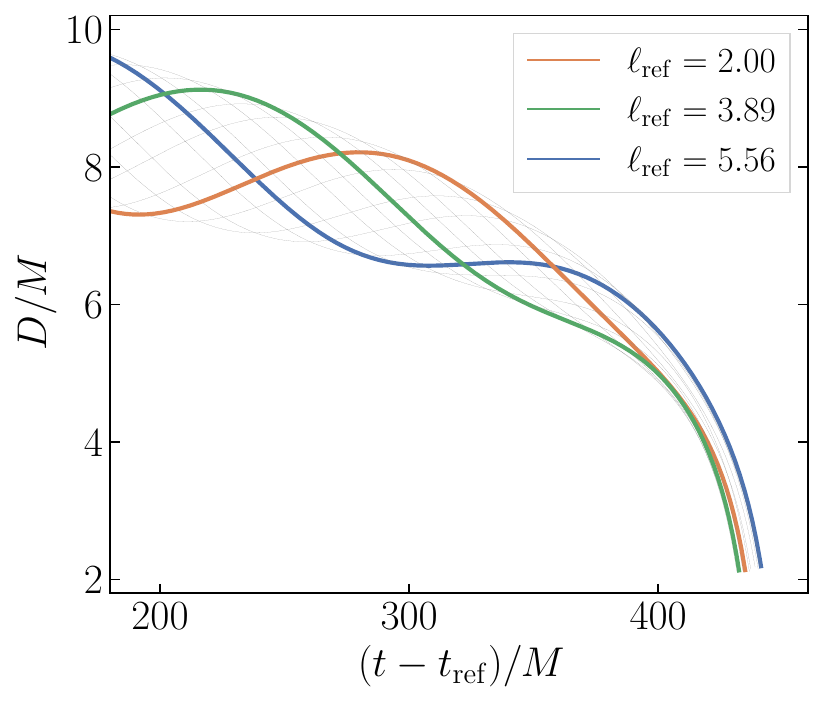}
    \caption{\textit{Sample sequence of 13 NR simulations spanning mean anomaly.} Each curve corresponds to a full NR simulation with $q=1$ and $\chi_i=0$, where the initial conditions were generated using our new eccentricity control method. The common envelope showcases the control obtained over \textit{both} eccentricity and mean anomaly. Three simulations are highlighted in color.
    }
    \label{plot: separation}
\end{figure*}

To isolate the impact of both eccentricity and mean anomaly, we must ensure that we are covering the entire two-dimensional $(e,\ell)$ parameter space in a controlled way. Previous works~\cite{Healy:2022wdn, Radia:2021hjs} generate one-parameter sequences of simulations with increasing eccentricity, with little control over mean anomaly or relied on randomness to generate simulations with different mean anomaly~\cite{Islam:2021mha}.

To ensure proper attention is given to both eccentricity parameters, we seek to generate sequences of simulations with \textit{equal eccentricity}, but with uniform coverage over reference mean anomaly. To do so, we will use the eccentricity control method developed in Sec.~\ref{Sec:Methods} to construct a sequence of 13 simulations such that $T_{\rm merger}^{\rm target}$ changes by one orbital period during the sequence, and such that $e_0^{\rm target}$ changes slightly, to keep eccentricity at a reference epoch constant. Because all our simulations start at apastron, the variation of $T_{\rm merger}^{\rm target}$ through one radial period causes mean anomaly at the reference point to cycle through a full $[0,2\pi]$ interval.
The exact values are found via root-finding on EOB initial parameters, such that the EOB waveform at the reference epoch has the desired constant eccentricity and varying mean anomaly. The results, listed in table~\ref{table: ecc sequence}, are then used in the eccentricity tuning procedure of Sec.~\ref{Sec:Methods}. We will then generate 3 more of these sequences with 12 simulations each for varying values of orbital eccentricity, allowing us to probe the impact of eccentricity itself.

\begin{table}[]
    \label{table: ecc sequence}
    \centering
    \parbox{0.9\textwidth}{
        \begin{tabular}{l|l|l|l|l|l|l|l}
            $e_0^{\rm target}$            & 0.5000   & \textbf{0.5000}   & 0.4984   & 0.4974   & 0.4964   & \textbf{0.4948}   & 0.4937   \\ \hline
            $T_{\rm merger}^{\rm target}$ & $11915M$ & $\mathbf{11884M}$ & $11798M$ & $11719M$ & $11642M$ & $\mathbf{11557M}$ & $11478M$ \\
        \end{tabular} \\[1em]
        \begin{tabular}{l|l|l|l|l|l|l}
            $e_0^{\rm target}$            & 0.4925   & 0.4912   & \textbf{0.4899}   & 0.4886   & 0.4877   & 0.4877   \\ \hline
            $T_{\rm merger}^{\rm target}$ & $11398M$ & $11318M$ & $\mathbf{11238M}$ & $11158M$ & $11084M$ & $11021M$ \\
        \end{tabular}
    }
    \caption{$e_0^{\rm target}$ and $T_{\rm merger}^{\rm target}$ used for the mean anomaly sequence presented in figure~\ref{plot: separation}. Bold faced entries represent the three coloured simulations presented throughout the paper.    }
\end{table}

\begin{figure}
    \centering
    \includegraphics[width=0.9\linewidth]{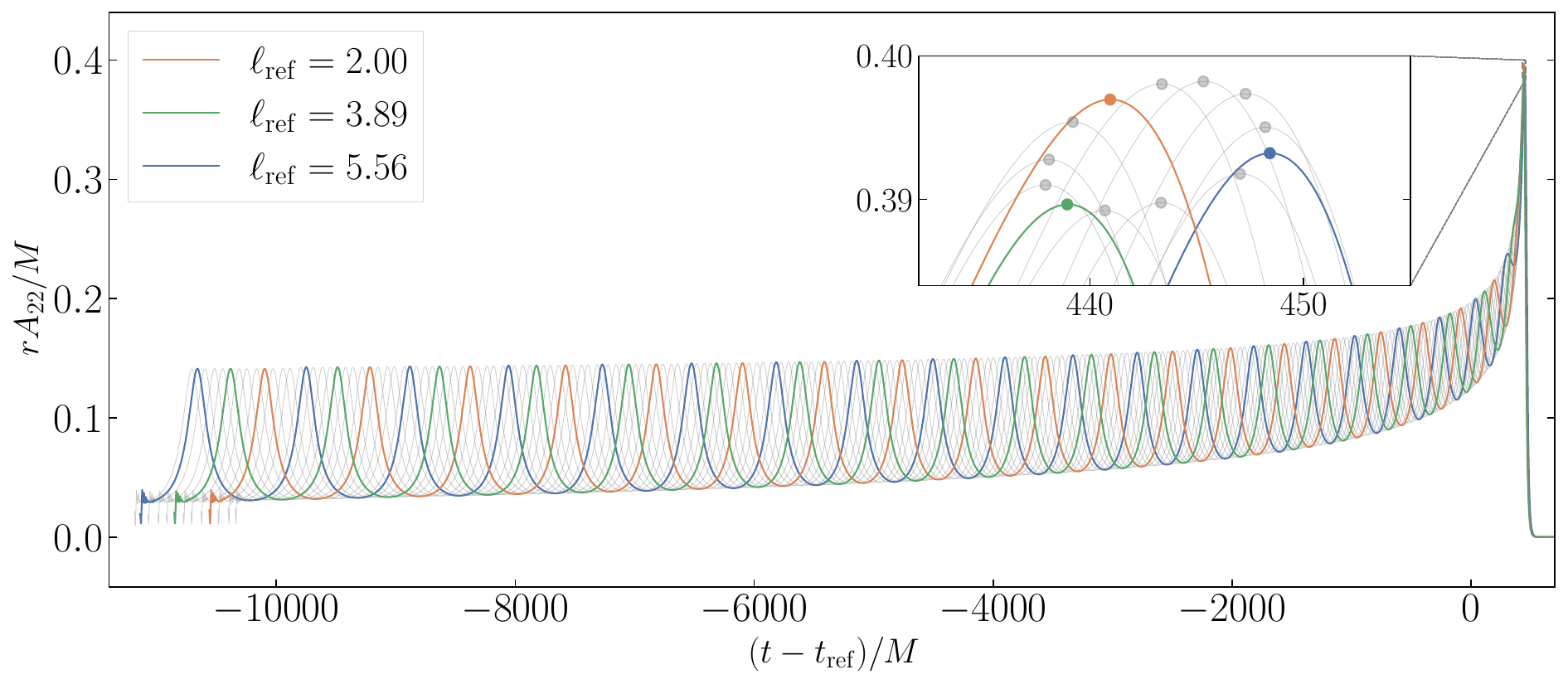}
    \caption{\textit{GW amplitude for the mean anomaly sequence with
            $e_{\rm ref}=0.17$}.  Shown is $A_{22}$ for the NR
        simulations arising from table~\ref{table: ecc sequence}.  Variations of mean anomaly across this sequence manifest themselves as ``sliding'' of the extrema during the inspiral, and as modulations of amplitude and time of $\max(rA_{22}/M)$, as indicated by the circles in the inset.}
    \label{plot: t_A22}
\end{figure}

Figure~\ref{plot: separation} showcases a sample sequence of 13 simulations of constant eccentricity.  Plotted is the coordinate separation
of the BHs throughout the simulation, with the time-axis
aligned at orbit-averaged frequency
$\langle\omega_{22}\rangle=0.069/M$.  We see
that all simulations in figure~\ref{plot: separation} share the same
envelope of maxima/minima of separation.  While we provide no proof,
it is expected that for most sensible evolution gauges, sequences of
equal eccentricity should share a common envelope of their respective
separation curves. We quantify how constant the eccentricity is within the sequence in the appendix~\ref{sec:appendix}. The right panel of figure~\ref{plot: separation}
zooms in on the transition-to-plunge portion of the simulation.
Depending on the specific reference mean anomaly, both the length and
shape of the final orbit can be markedly altered. This highlights that
the type of ``merger geometry'' obtained will depend on the specific
mean anomaly of each system.

Figure~\ref{plot: t_A22} plots the amplitude of the (2,2) mode $A_{22}$ for each of the simulations in table~\ref{table: ecc sequence}, aligned by the time at which the orbit-averaged frequency is $\langle\omega_{22}\rangle=0.069/M$. The inset shows a zoom-in on the merger portion of the waveform. Again, the shared envelope formed by the local maxima of $A_{22}$ demonstrates the constant reference eccentricity throughout the sequence. We observe that both the amplitude of the peak of the $(2,2)$ mode, as well as the time taken from $t_{\rm ref}$ to the peak show a cyclic dependence on the reference mean anomaly.

\begin{figure*}
    \includegraphics[width=0.5\linewidth]{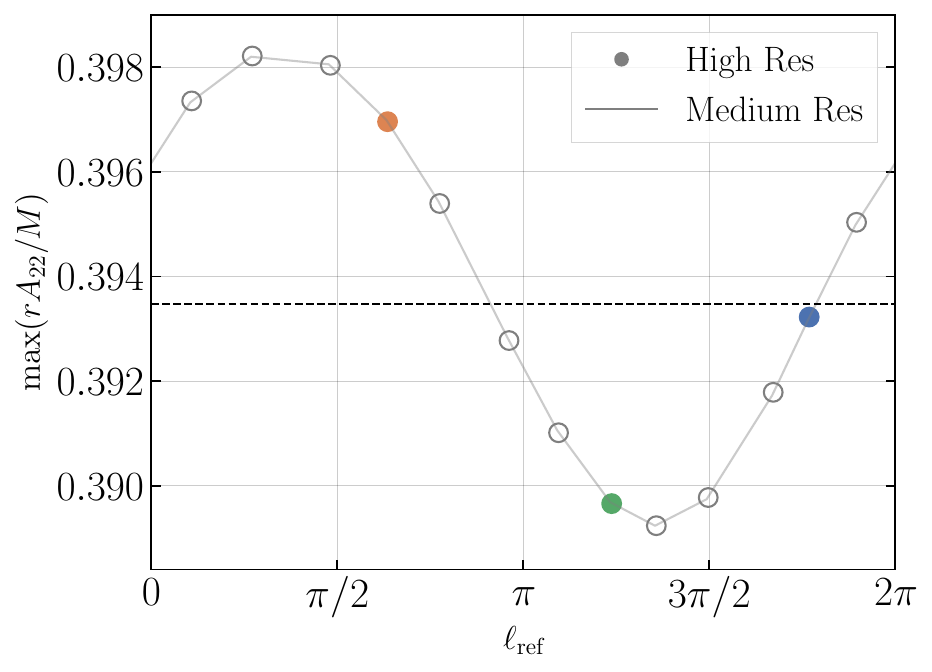}
    \includegraphics[width=0.5\linewidth]{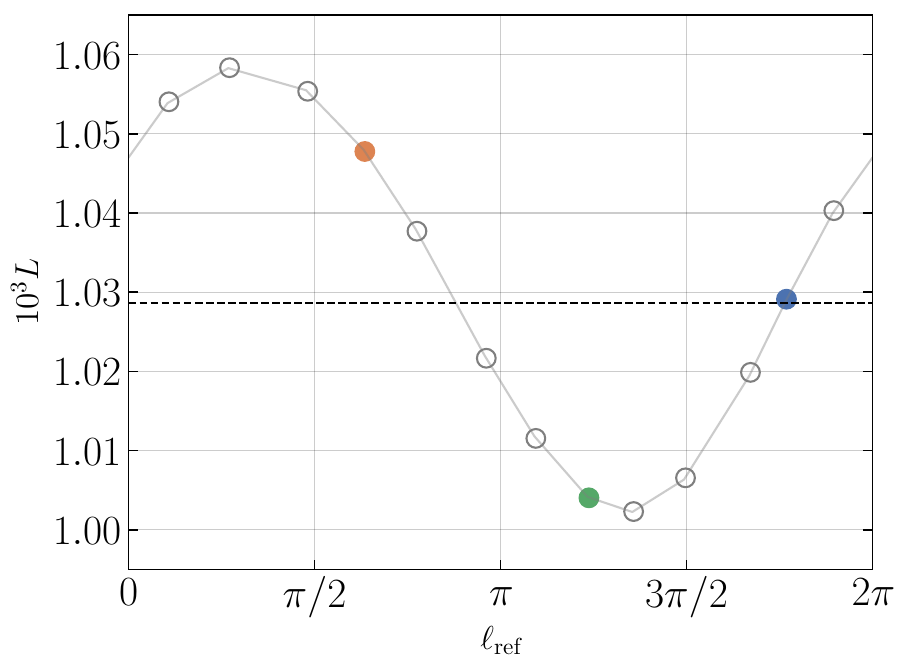}
    \includegraphics[width=0.5\linewidth]{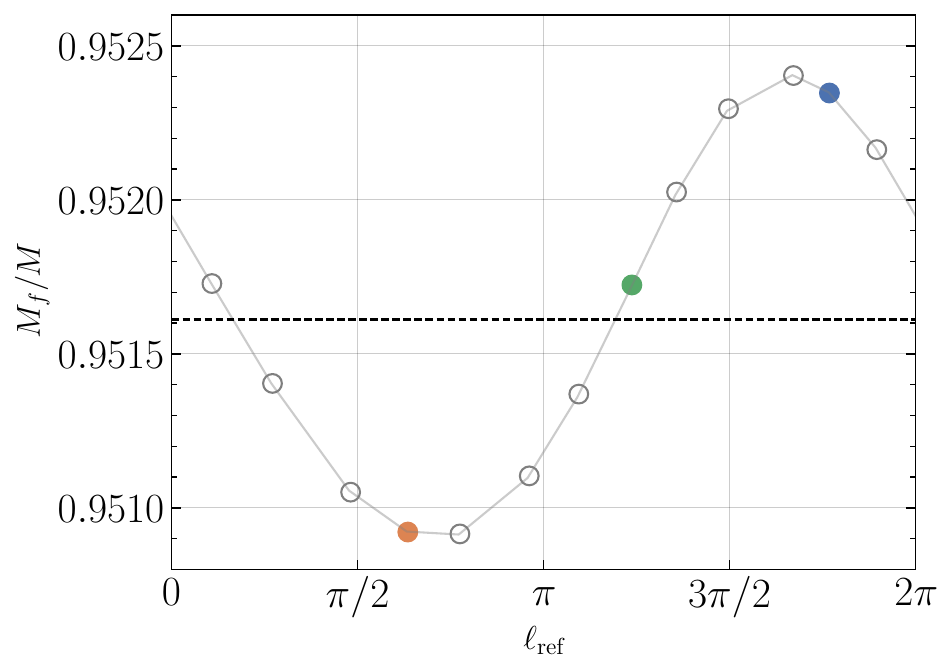}
    \includegraphics[width=0.5\linewidth]{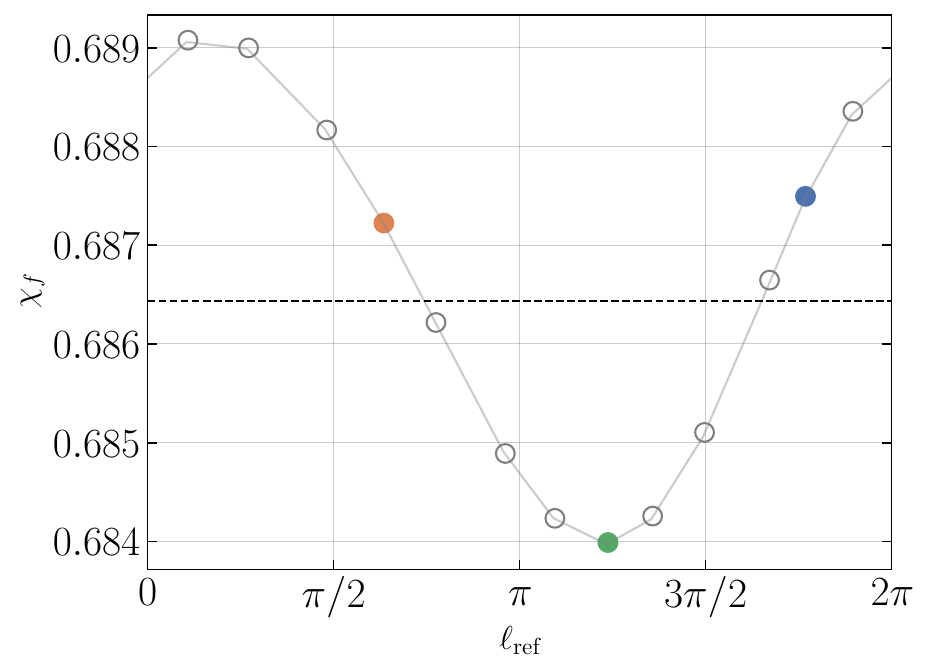}
    \caption{\textit{Dependence of merger quantities on mean anomaly.} Here, we plot the peak of $A_{22}$, peak luminosity, and remnant mass and spin for the systems shown in figure~\ref{plot: separation}. Note that each quantity can be both increased and decreased relative to their quasi-circular value (indicated by the horizontal dashed black lines). Colored points correspond to the highlighted simulations in figure~\ref{plot: separation}.  The faint lines in each panel represent results from lower-resolution numerical simulations; they are visually indistinguishable from the high-resolution simulations plotted as symbols, showcasing that numerical truncation error is subdominant.
    }
    \label{plot: mean ano dependence}
\end{figure*}

Let us now explore the dependence on mean anomaly within this sequence
of simulations in more detail, continuing to use
a reference frequency $\langle\omega_{22}\rangle=0.069/M$ for
reporting mean anomaly and eccentricity.  Specifically, we consider the
maximum of the amplitude of the $(2,2)$ GW-mode ($\max(rA_{22}/M)$), the
peak GW luminosity ($L$) and the remnant mass and remnant spin ($M_f$
and $\chi_f$).
Figure~\ref{plot: mean ano dependence} plots
these four quantities as a function of the reference mean anomaly.

We
see that the dependence of each of these quantities on mean anomaly is
oscillatory, with the largest deviations (in $L$) having relative magnitudes of $\sim5\%$. The horizontal dashed lines in each panel represent the corresponding value for quasi-circular inspirals demonstrating that eccentricity can lead to either an increase
or decrease in merger quantities, depending on the specific value of
mean anomaly obtained.

Given our parametrization, we expect the BH systems to be $2\pi$-periodic in mean anomaly, and to vary smoothly with respect to mean anomaly. As such, we can expand the dependence of any particular quantity in terms of a Fourier series in mean anomaly.
The fact that the oscillations are (nearly) about the quasi-circular value indicates that this is the dominant contribution to the constant term in the expansion, and figure~\ref{plot: mean ano dependence} indicates that  the first mode of the Fourier series dominates the oscillatory behavior.
We expand the data plotted in figure~\ref{plot: mean ano dependence} in a series of the form
\begin{equation}\label{eq: expansion}
    \delta X\equiv X-X_{\rm QC}  = A_0 + \sum_{k=1}^mA_k\sin(k\ell+\phi_k).
\end{equation}
Here $X$ denotes any of our four quantities of interest, and we subtract off the quasi-circular value so that $A_0$ directly represents the mean anomaly averaged change in that value due to eccentricity.
We keep the first three terms in the expansion, $m=2$. While we can fit for higher harmonics, we find that our limited dataset makes it difficult to ascertain the accuracy of these fits.

\begin{figure*}
    \includegraphics[width=0.5\linewidth]{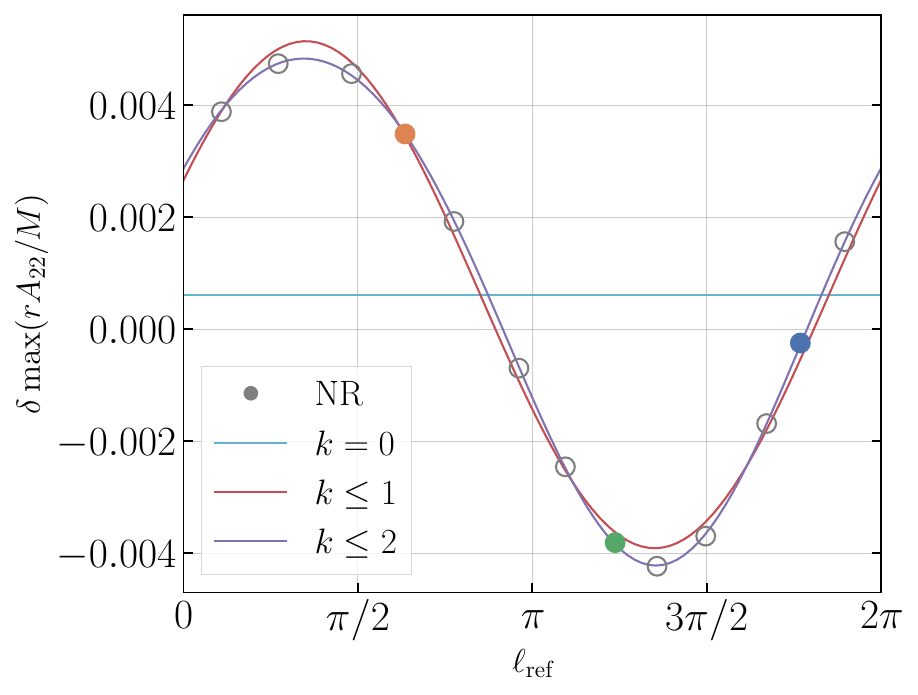}
    \includegraphics[width=0.5\linewidth]{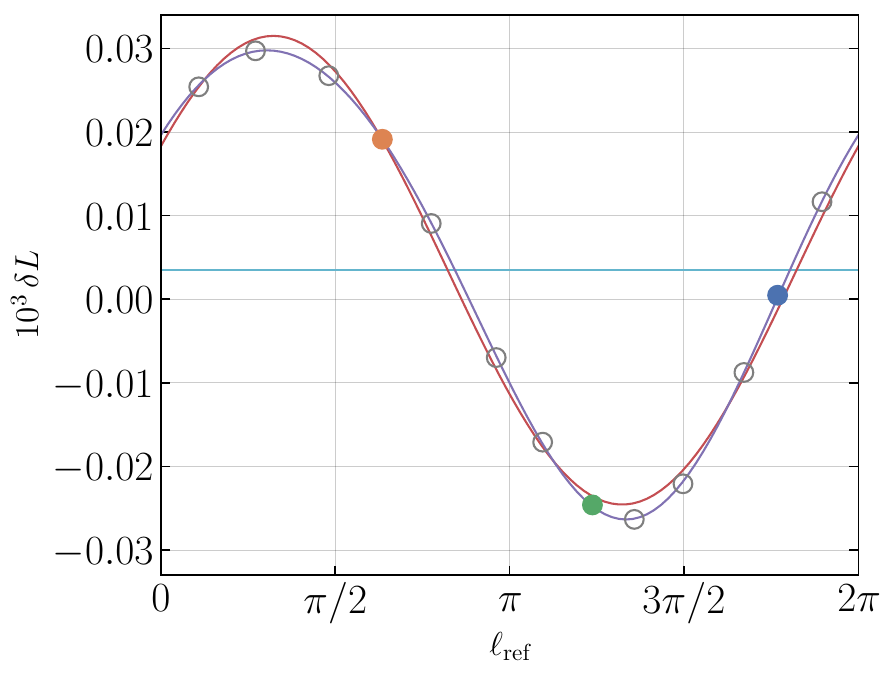}
    \includegraphics[width=0.5\linewidth]{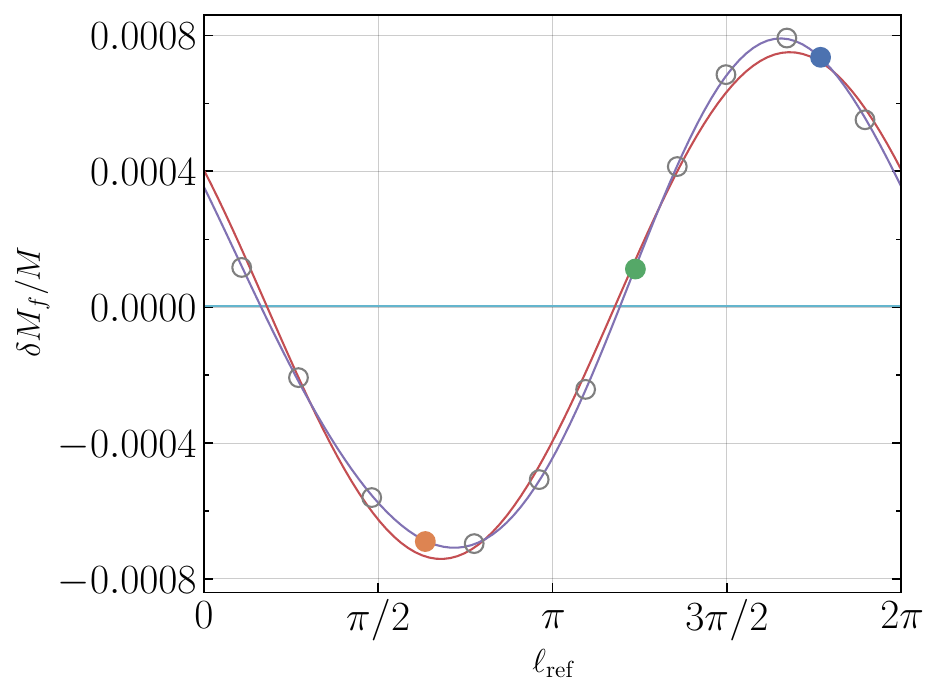}
    \includegraphics[width=0.5\linewidth]{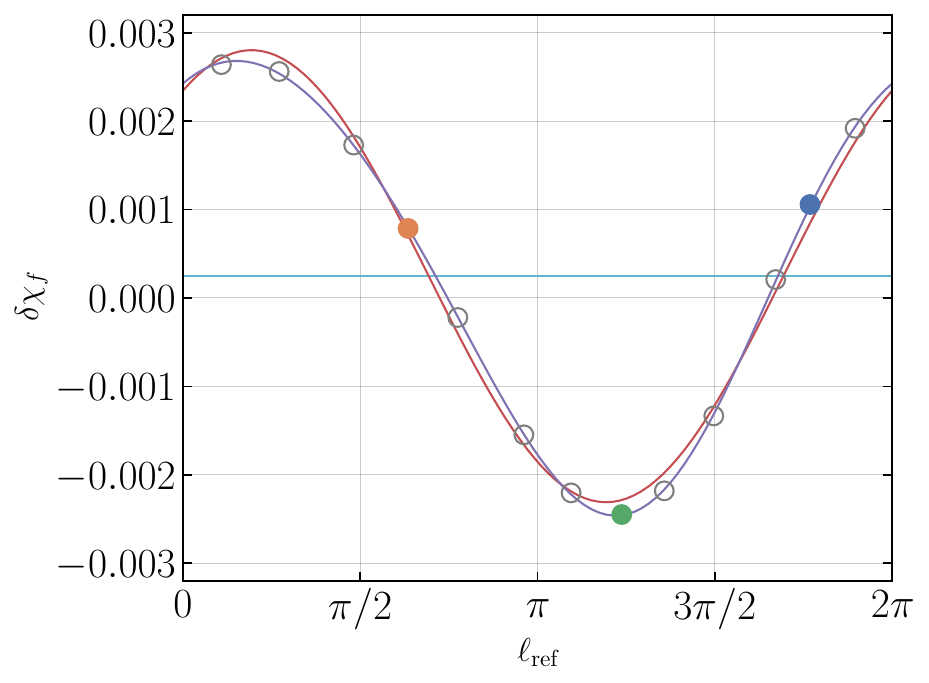}
    \caption{\textit{Presence of different harmonics in each quantity.} Plotted are partial sums of~(\ref{eq: expansion}), including only the constant term ($k=0$), up to first harmonic ($k\leq 1$) and up to second harmonic ($k\leq 2$).
    }
    \label{plot: harmonics}
\end{figure*}

Figure~\ref{plot: harmonics} plots partial sums of the
expansions~(\ref{eq: expansion}), confirming that the constant term is slightly different
from the quasi-circular value, and that the first mode dominates the
oscillations.  The second harmonic is a factor $\sim 10$ smaller
than the first harmonic, still large enough to be easily visible in
figure~\ref{plot: harmonics}.

\begin{figure}
    \centering
    \includegraphics[width=0.7\linewidth]{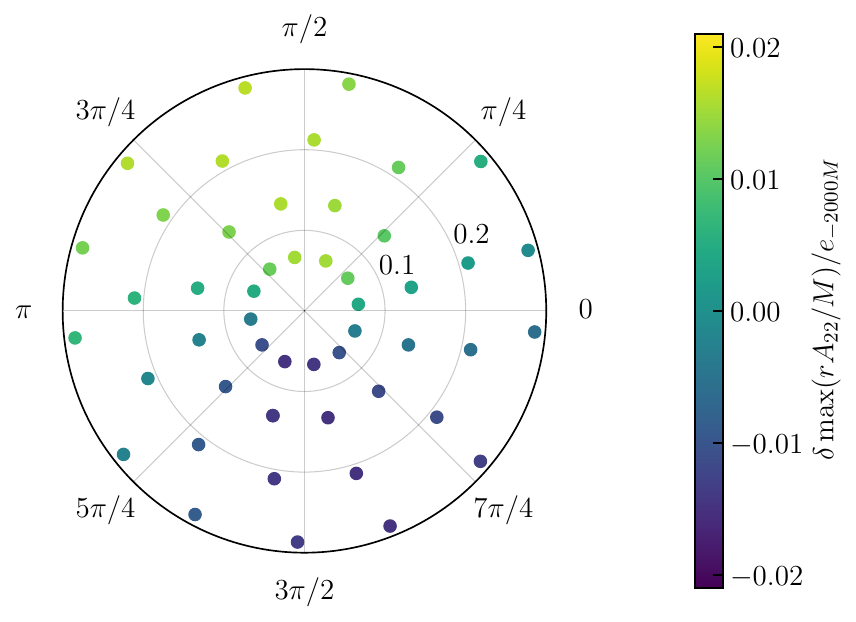}
    \caption{\textit{Deviation of $\max(rA_{22}/M)$ across eccentricities.} Plotted are 4 sequences of NR simulations of varying eccentricity (indicated by constant radius rings), with eccentricities $e_{-2000M}=0.095, 0.185, 0.272, 0.346$. The angle of each point corresponds to $\ell_{-700M}$, while distance from the origin is equal to $e_{-2000M}$. Each point is coloured by the deviation of the peak of $A_{22}$ divided by $e_{-2000M}$, to highlight that the range of $\delta \max(rA_{22}/M)$ grows approximately linearly in $e$ (as each ring spans the same $\delta \max(rA_{22}/M)/e_{-2000M}$ range).}
    \label{plot: A22_polar}
\end{figure}

Turning towards the dependence on eccentricity, we choose three additional values of reference eccentricity, and perform NR simulations at several values of reference mean anomaly, bringing our set of simulations up to 49 simulations.
Figure~\ref{plot: A22_polar} represents $(e,\ell)$ for these simulations in a polar plot, highlighting the uniform coverage
our technique achieves.
The color-coding of figure~\ref{plot: A22_polar} conveys information about the peak-amplitude of the $(2,2)$ GW mode for all the simulations.  Specifically, the color represents the difference between $\max(rA_{22}/M)$ of each eccentric simulation to that of the quasi-circular simulation, normalized by each simulation's eccentricity.
Going around the center of figure~\ref{plot: A22_polar} at fixed eccentricity, one sees one oscillation of $\delta \max(rA_{22}/M)$, duplicating the oscillatory behavior of $\max(rA_{22}/M)$ seen in figure~\ref{plot: mean ano dependence}.
Going radially in figure~\ref{plot: A22_polar}, one notices that the colors are nearly constant; this indicates that $\delta \max(rA_{22}/M)$ is proportional to the eccentricity of the simulations.

The data plotted in Figs.~\ref{plot: mean ano dependence} and
\ref{plot: A22_polar} depends on the reference point chosen to extract
eccentricity and mean anomaly.  If the reference point moves earlier
(farther before merger), then the extracted eccentricities will be
larger, owing to the decay of eccentricity during the inspiral.
Furthermore, the extracted mean anomalies will cycle once through the
interval $[0, 2\pi]$ with each radial oscillation period the reference
point moves earlier. Because the inspiral rate of eccentric binaries
depends on their eccentricity, the mean anomaly values extracted at
earlier times will dephase between simulations with different
eccentricities: The very simple overall behavior seen in
figure~\ref{plot: A22_polar} is only present when $\ell$ is extracted
sufficiently close to merger (in the figure, at $t-t_{\rm
    peak}=-700M$).  Since it is the merger behavior that determines
$\delta \max(rA_{22}/M)$, it comes at no surprise that its dependence on
$(e,\ell)$ is simplest when a reference point very near merger is
chosen.

\begin{figure}
    \centering
    \includegraphics[width=\linewidth]{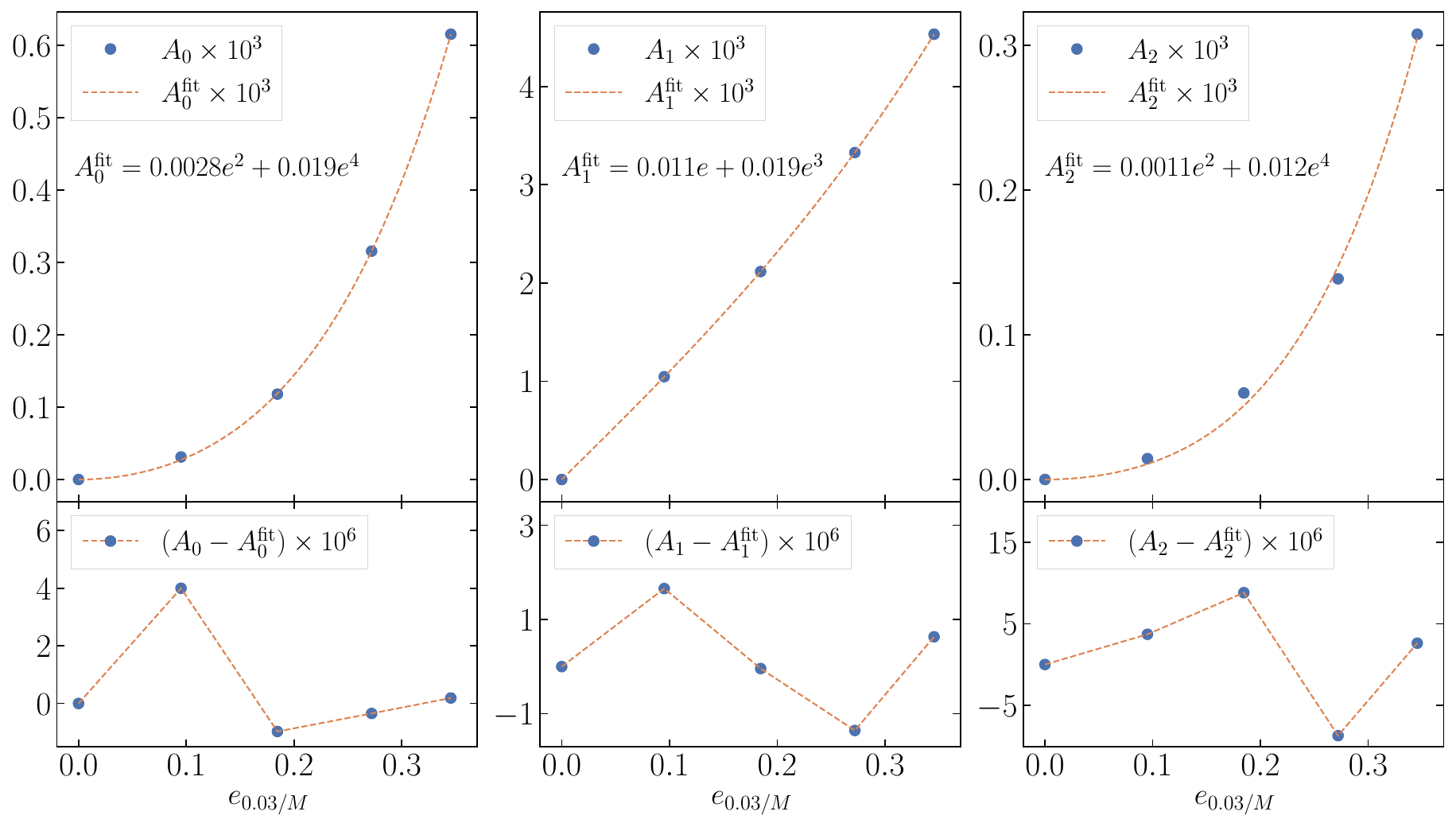}
    \caption{\textit{Scaling of mean anomaly harmonic amplitudes with eccentricity.} We perform the analysis of figure~\ref{plot: harmonics} for each of the constant eccentricity sequences shown in figure~\ref{plot: A22_polar}. Plotted are the amplitudes $A_{i}$ of the expansion of peak GW amplitude $A_{22}$ in the form of~(\ref{eq: expansion}). The dashed lines correspond to fits of~(\ref{eq: even}) and (\ref{eq: odd}) as appropriate.  We include an additional data point for $e_{0.03/M}=0$, as we know the correction to the quasi-circular value must vanish. The lower panel shows the residuals of the fits.}
    \label{plot: ecc dependence}
\end{figure}

While figure~\ref{plot: A22_polar} indicates that the overall scale of the oscillations scales linearly with eccentricity, one might wonder how the individual harmonics presented in figure~\ref{plot: harmonics} scale with increasing eccentricity. Focusing on $\delta\max(rA_{22}/M)$, we repeat fits of the form~(\ref{eq: expansion}) separately for the four different mean anomaly sequences at different eccentricities,
and plot the obtained amplitudes in the top panels of
figure~\ref{plot: ecc dependence}. Figure~\ref{plot: harmonics} indicates that the dominant contribution to~(\ref{eq: expansion}) is the first harmonic $A_1$, which we find indeed scales almost linearly.

To understand the phenomonelogy of these scalings, first note that the amplitudes of~(\ref{eq: expansion}) can be extended to depend on eccentricity using a low eccentricity expansion:

\begin{equation}\label{eq: ecc expansion}
    A_{k}(e) = a_1e +a_2e^2 + a_3e^3+a_4e^4+....
\end{equation}

Because we are expanding the difference to the quasi-circular limit, the constant $a_0$ term vanishes. Consider a system characterised by a reference $(e,\ell)$. Under the inter-change of the definition of periastron and apastron, we observe that the same physical system would instead be characterised by $(-e,\ell+\pi)$. As such, any eccentric correction should be invariant under this re-definition. This limits the number of possible terms in the eccentric expansion of $A_k$, as these terms come with a factor of $\sin(k\ell)$. Specifically, keeping only the first two terms in the expansion, for even $k$ we should expect the dependence on $e$ to be
\begin{equation}\label{eq: even}
    A_k=a_2e^2 + a_4e^4,
\end{equation}
while for odd $k$, we obtain
\begin{equation}\label{eq: odd}
    A_k=a_1e + a_3e^3.
\end{equation}

The dashed curves in figure~\ref{plot: ecc dependence} show fits of~(\ref{eq: even}) and (\ref{eq: odd}) as appropriate. We find that these fits match the data well, approaching the truncation error of our data. However, a more careful consideration of these fits (including the exploration of higher-order contributions) will be required to build fits of these deviations across parameter space.

\subsection{Interpreting 1-parameter eccentric sequences}\label{sec:32}

With our comprehensive NR data-set in hand, and the accompanying understanding of eccentric binaries, we
can now discuss the phenomenology observed in other eccentric explorations. Several other works have reported a seemingly oscillatory dependence of merger-related quantities (e.g., maximum waveform strain $\max(rA_{22}/M)$) on eccentricity \cite{Wang:2023vka, Wang:2023wol, Wang:2024jro,Radia:2021hjs, Carullo:2023kvj}. In contrast, figure~\ref{plot: ecc dependence} indicates a monotonic dependence on eccentricity.
Our claim is that such reported oscillations arise from taking a \textit{one-parameter} sequence of simulations across the two-dimensional $(e,\ell)$ parameter space.  As $e$ changes along such a one-parameter sequence, typically $\ell$ will also change in some way, related to the precise description being used to construct the one-parameter sequence. It is then the dependence on $\ell$ (as exhibited in figure~\ref{plot: mean ano dependence}) that induces a seemingly ``oscillatory behavior with eccentricity''.

\begin{figure}
    \centering
    \includegraphics[width=0.8\linewidth]{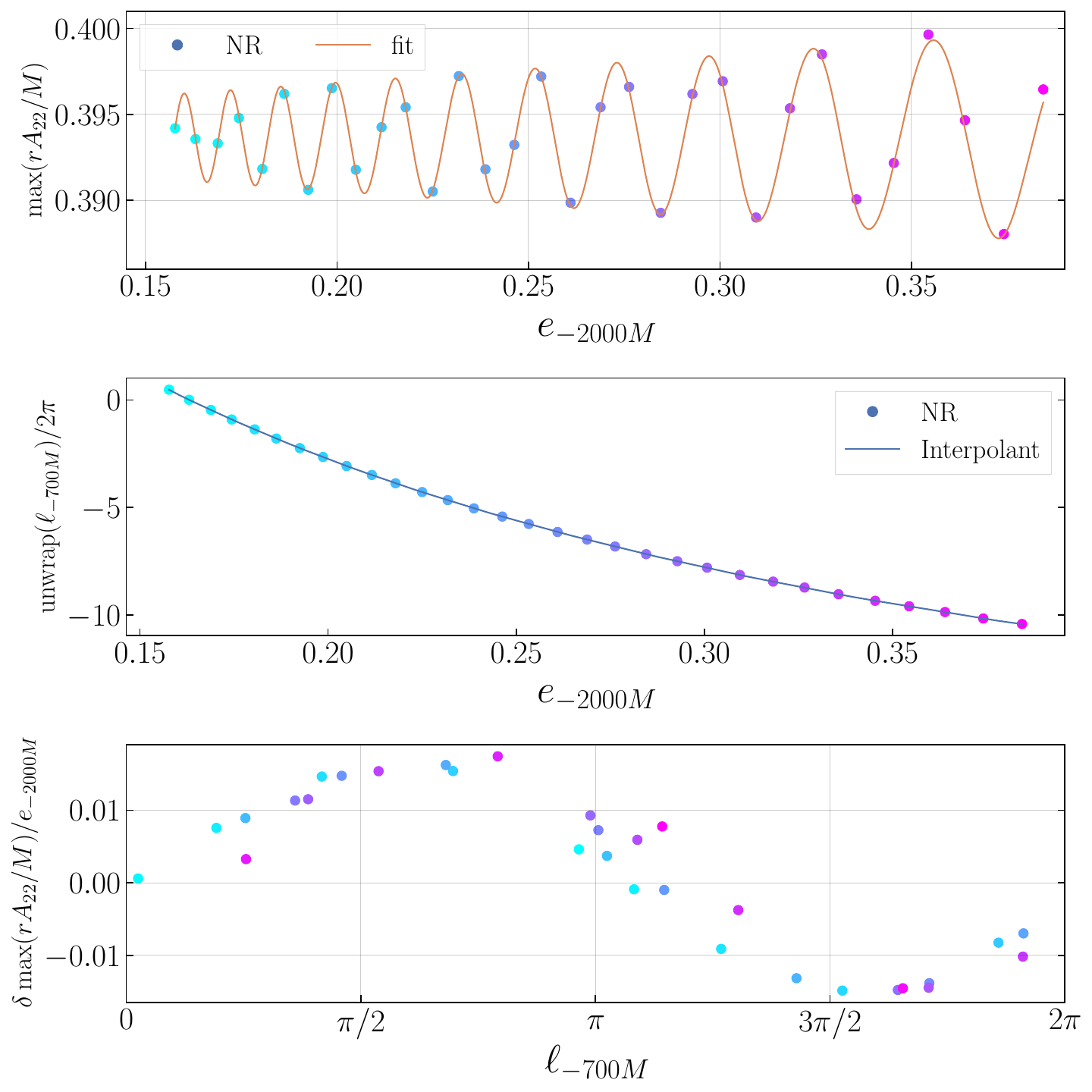}
    \caption{
        \textit{Typical sequence obtained if mean anomaly is neglected.} Sequence consists of simulations where the initial separation and in-going velocity of the BHs is fixed, but the initial orbital frequency $\Omega_0$ is slowly decreased. \textbf{Top:} Peak of $A_{22}$ for the sequence. The orange line corresponds to the model~(\ref{eq:A22-sequence}). \textbf{Middle:} Unwrapped reference mean anomaly as a function of the reference eccentricity. \textbf{Bottom:} Deviation of the peak of $rA_{22}/M$ as a function of the mean anomaly. Coloring corresponds to $e_{-2000M}$.}
    \label{plot: sequence}
\end{figure}

To support our claim, we generate a one-dimensional sequence of eccentric BBH simulations following
the procedure of Healy et al.~\cite{Healy:2022wdn}:
we fix the intrinsic BH parameters to be non-spinning, equal mass
throughout this calculation, and begin with $(D_0, \Omega_0,
    \dot{a}_0)$ such that the BHs are in a quasi-circular
configuration.  We choose a large initial separation of $D_0\approx24.1M$,
corresponding to a quasi-circular inspiral of duration $30000M$, in
order to always have a sufficiently long inspiral to measure
eccentricity and mean anomaly with \texttt{gw\_eccentricity}. This
furthermore will allow us to exhibit a large number of oscillations of
$\max(rA_{22}/M)$ as we proceed through this sequence.  We then generate a
sequence of simulations where we keep $D_0$ and $\dot{a}_0$ unchanged,
but decrease $\Omega_0$, resulting in simulations with
increasing initial eccentricity and decreasing time to merger.  For
each simulation, we estimate eccentricity at a reference time
$2000M$ before merger, and we record the maximum amplitude of the
$(2,2)$ mode of the emitted gravitational radiation, $\max(rA_{22}/M)$.
Results from 31 NR simulations are plotted as filled circles in the
top panel of figure~\ref{plot: sequence}.  This plot of $\max(rA_{22}/M)$ vs
$e_{-2000M}$ indeed appears to exhibit oscillatory behavior with
eccentricity. However, the simulations of this sequence will also
have different mean anomaly at a fixed reference epoch, here chosen as
$700M$ before merger.  In fact, each time the number of radial periods
of the inspiral decreases by one, we should approximately cycle
through a full period of reference mean anomaly near merger.  The
mean anomaly of each of the 31 simulations is plotted in the middle
panel of figure~\ref{plot: sequence}, made continuous by suitable additions
of $2\pi$, and divided by $2\pi$, so that the y-axis represents the
difference in number of radial periods each simulation completes,
relative to the simulation with the smallest eccentricity.

To interpret the NR data shown in figure~\ref{plot: sequence}, we utilize the functional form for the peak amplitude that we derived in Sec.~\ref{sec:31}, namely
\begin{equation}\label{eq:A22-sequence}
    \max(rA_{22}/M)(e_{-2000M},\ell_{-700M}) = \max(rA_{22}/M)^{\rm QC} + e_{-2000M} A_1 \sin(\ell_{-700M} +\ell_0),
\end{equation}
where for simplicity we have elected to keep only the first harmonic in mean anomaly (as well as the first term in the eccentricity expansion), and exclude the constant offset.
Here $\ell_0$
represents the offset in mean anomaly at reference epoch to the
phasing of the merger amplitude $\max(rA_{22}/M)$.  The 1-parameter
sequence of simulations covers a wide range of eccentricities, with
correspondingly somewhat different inspiral rates; therefore, $\ell_0$
will mildly depend on eccentricity, even for our reference epoch
$t=-700M$ close to merger\footnote{Here, as in figure~\ref{plot: A22_polar}, we choose to define eccentricity and mean anomaly at different times to merger. Currently, \texttt{gw\_eccentricity} produces oscillatory eccentricities very close to merger, likely due to the spline interpolation. To avoid this, we define our reference eccentricity earlier in the inspiral.}.  Because the variations in $\ell_0$ are
mild, we approximate it as a first order Taylor expansion,
\begin{equation}\label{eq:ell_0-Taylored}
    \ell_0(e_{-2000M}) =B_0 + B_1 e_{-2000M}.
\end{equation}
Our NR simulations give us 31 data-points for $\max A_{22}$ in the \textit{two-}dimensional $(e,\ell)$ plane, and~(\ref{eq:A22-sequence}) with~(\ref{eq:ell_0-Taylored}) represents the expected behavior of this quantity, given the three parameters $A_1, B_0, B_1$.  We now fit these three unknown parameters to the NR data, and thus achieve the full functional form $\max(rA_{22}/M)(e,\ell)$.  The result of this fit is plotted as a solid curve in the top panel of figure~\ref{plot: sequence},
where we evaluate $\ell_{-700M}(e_{-2000M})$ for arbitrary $e_{-2000M} $ via the interpolant shown in blue in the middle panel.
The three-parameter fit reproduces all variations in the data-points, exhibiting that indeed within the eccentricity range considered in figure~\ref{plot: sequence}, the 1-dimensional sequence of simulations proceeds through 10 maxima and minima of $\max(rA_{22}/M)$, corresponding to 10 cycles through mean anomaly.
Finally, the bottom panel of figure~\ref{plot: sequence} plots the
deviations of $\max(rA_{22}/M)$ from the quasi-circular value as a function of mean anomaly.
After normalization by eccentricity, the data approximately collapses to a single sinusoidal curve, reminiscent of figure~\ref{plot: mean ano dependence}.

Compared to other works \cite{Wang:2023vka, Wang:2023wol, Wang:2024jro}, our dataset seemingly contains many more oscillations. This arises from the length of our simulations: as previously stated, each time the number of radial periods decreases by one, the mean anomaly defined near merger will cycle through $2\pi$. One conclusion is that even for significant eccentricities near merger, one can approximate the behavior of several quantities in the two-dimensional $(e,\ell)$ space with simple formulae (requiring only 3 parameters fit from data).

\subsection{Effective-one-body estimates of the final mass and spin}

While several of our results may only be accessible via full NR simulations, there are hints that others, such as the final mass and spin of the remnant BH, may be well approximated using semi-analytical methods. The idea of analytically estimating the final mass and spin of a BBH coalescence was introduced in Ref.~\cite{Buonanno:2000ef, Buonanno:2005xu}. In this work, the final mass and spin were estimated from the EOB Hamiltonian $H_{\rm{EOB}}$ and orbital angular momentum $p_{\phi}$ at the end of the plunge, providing a prediction of such quantities for a remnant generated by a BBH merger.
With the breakthrough of NR~\cite{Pretorius:2005gq,Campanelli:2005dd,Baker:2005vv}, the picture outlined in Ref.~\cite{Buonanno:2000ef} was shown to be qualitatively, and to some extent quantitatively, correct~\cite{Buonanno:2006ui}.
The main physical ingredient which is lacking in these early estimates is the loss of energy and angular momentum during the ringdown phase. Later works~\cite{Damour:2007cb,Damour:2013tla} refined the accuracy of EOB estimates by approximately accounting for such losses, as
\begin{align}
    \label{eq:final_mass_eob}
    M_f & = H_{\rm{EOB}}(t_{\rm{match}}) - E_{\rm{ringdown}}, \\
    J_f & = J_{\rm{EOB}}(t_{\rm{match}}) - J_{\rm{ringdown}}.
\end{align}
Here, $t_{\rm{match}}$ represents the time at which the inspiral-plunge and ringdown waveforms are connected, around the peak of the EOB orbital frequency. The energy and angular momentum losses during the ringdown, $E_{\rm{ringdown}}$ and $J_{\rm{ringdown}}$, were approximated by a suitable rescaling of test-particle results~\cite{Nagar:2006xv}.
Recent EOB models compute these quantities from more accurate NR fits for binaries on quasi-circular orbits~\cite{Jimenez-Forteza:2016oae, Hofmann:2016yih}.
While analogous fits have started to become available for eccentric BBHs~\cite{Carullo:2023kvj}, the limited number of NR simulations covering the eccentric parameter space prevents a precise assessment of their accuracy, particularly when spin effects are also included.

Here, we present an approximate method which offers a practical approach to account for merger effects in current EOB models for eccentric BBHs, relying only on quasi-circular NR fits.
Focusing on the final mass, let us consider~\eqref{eq:final_mass_eob} for an eccentric and a quasi-circular BBH with the same component masses and spins,
\begin{align}
    M_f^{\rm{ecc}} & =H_{\rm{EOB}}(t_{\rm{match}}^{\rm{ecc}}) - E^{\rm{ecc}}_{\rm{ringdown}}, \\
    M_f^{\rm{QC}}  & =H_{\rm{EOB}}(t_{\rm{match}}^{\rm{QC}}) - E^{\rm{QC}}_{\rm{ringdown}}.
\end{align}
Since the binary circularizes during the inspiral, we assume that the energy emitted during the ringdown stage is similar in the two cases,
\begin{equation}
    E^{\rm{ecc}}_{\rm{ringdown}}\simeq E^{\rm{QC}}_{\rm{ringdown}}.
\end{equation}
We can then approximate the final mass for an eccentric BBH as
\begin{equation}
    \label{eq:final_mass_eob_ecc}
    M_f^{\rm{ecc}}\simeq M_f^{\rm{QC}} + \left[ H_{\rm{EOB}}(t_{\rm{match}}^{\rm{ecc}}) - H_{\rm{EOB}}(t_{\rm{match}}^{\rm{QC}}) \right].
\end{equation}
In this equation, the $M_f^{\rm{QC}}$ value can be obtained from NR fits for quasi-circular BBH mergers,
or ---for the present study of equal-mass, non-spinning BBH--- it can be directly read off from NR simulations~\cite{Scheel:2008rj}. The second term can be estimated using the \texttt{SEOBNRv5EHM} eccentric waveform model \cite{Gamboa:2024hli}.
In this model, the value of $t_{\rm{match}}^{\rm{QC}}$ is calibrated to quasi-circular BBH NR simulations, while $t_{\rm{match}}^{\rm{ecc}}$ is not calibrated to eccentric NR simulations.
However, the predicted attachment time $t_{\rm{match}}^{\rm{ecc}}$ employs indirectly the calibration to quasi-circular NR simulations from the \texttt{SEOBNRv5HM} model \cite{Pompili:2023tna}, hence making~\eqref{eq:final_mass_eob_ecc} a reasonable approximation.\footnote{A similar argument was applied in Ref.~\cite{Julie:2024fwy} to estimate corrections for the final mass and spin in Einstein-scalar-Gauss-Bonnet gravity relative to their value in General Relativity.}

Analogously, the final spin of an eccentric BBH merger can be estimated via
\begin{align}
    \begin{aligned}
        \label{eq:final_spin_eob_ecc}
        J_f^{\rm{ecc}}    & \simeq J_f^{\rm{QC}} + \left[ p_{\phi}(t_{\rm{match}}^{\rm{ecc}}) - p_{\phi}(t_{\rm{match}}^{\rm{QC}}) \right], \\
        \chi_f^{\rm{ecc}} & = J_f^{\rm{ecc}} / (M_f^{\rm{ecc}})^2,
    \end{aligned}
\end{align}
where $J_f^{\rm{QC}}$ can also be obtained from NR fits for quasi-circular BBHs. In the rest of this section, we use the \texttt{NRSur7dq4Remnant} fits~\cite{Varma:2019csw} for $M_f^{\rm{QC}}$ and $J_f^{\rm{QC}}$.

To assess the accuracy of the estimates Eqs.~(\ref{eq:final_mass_eob_ecc}) and~(\ref{eq:final_spin_eob_ecc}), we need a mapping between eccentric EOB waveforms and the NR results presented in the previous section, which is complicated because many definitions (including the ones employed in many waveform models) of eccentricity are gauge dependent~\cite{Ramos-Buades:2021adz,Knee:2022hth,Bonino:2024xrv}.
For the purpose of this work, we employ the method described in Ref.~\cite{Gamboa:2024hli} to find the optimum EOB waveform given an eccentric NR simulation.
This method is based on the one presented in Ref.~\cite{Ramos-Buades:2021adz}, and consists of optimizing over the EOB input values of eccentricity and orbit-averaged frequency to get the waveform with the lowest $ (2,2) $-mode unfaithfulness; in this optimization, the waveforms are initialized at apastron.
Given the high NR-faithfulness of the \texttt{SEOBNRv5EHM} model~\cite{Gamboa:2024hli}, this approach leads to a good agreement between the gauge-invariant values of eccentricity $ e _{ \mathrm{gw}} $ and mean anomaly $ \ell_{ \mathrm{gw}} $ extracted from the EOB and NR waveforms at the same reference frequency.
In this way, for each NR simulation, we calculate the values of $ M_f^{\rm{ecc}} $ and $ \chi_f^{\rm{ecc}} $ from Eqs.~\eqref{eq:final_mass_eob_ecc} and \eqref{eq:final_spin_eob_ecc} using the \texttt{SEOBNRv5EHM} model.

\begin{figure*}
    \includegraphics[width=0.52\linewidth,trim=8 8 0 7]{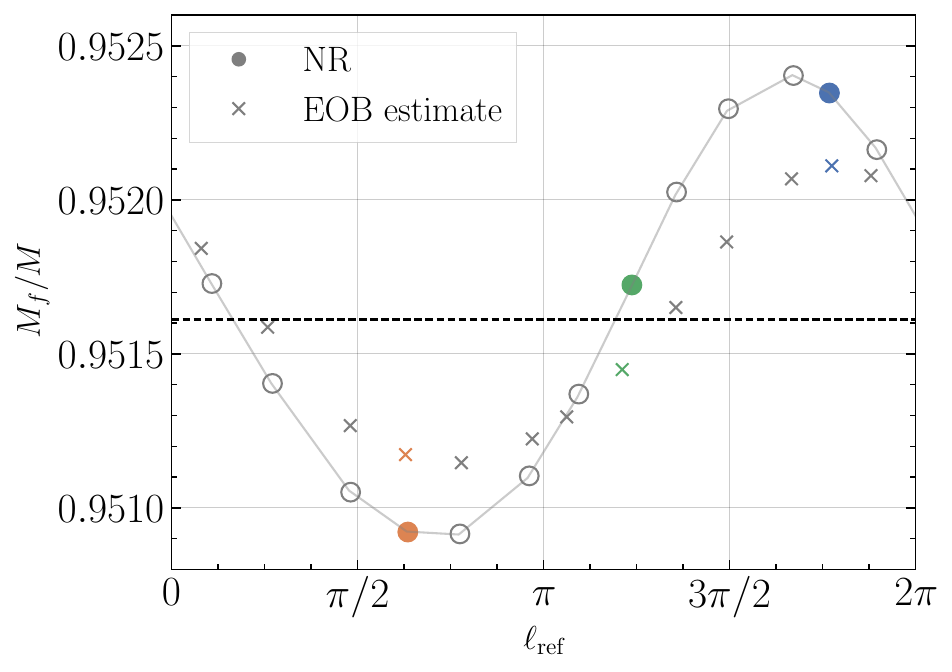}
    \includegraphics[width=0.47\linewidth,trim=0 -17 0 7]{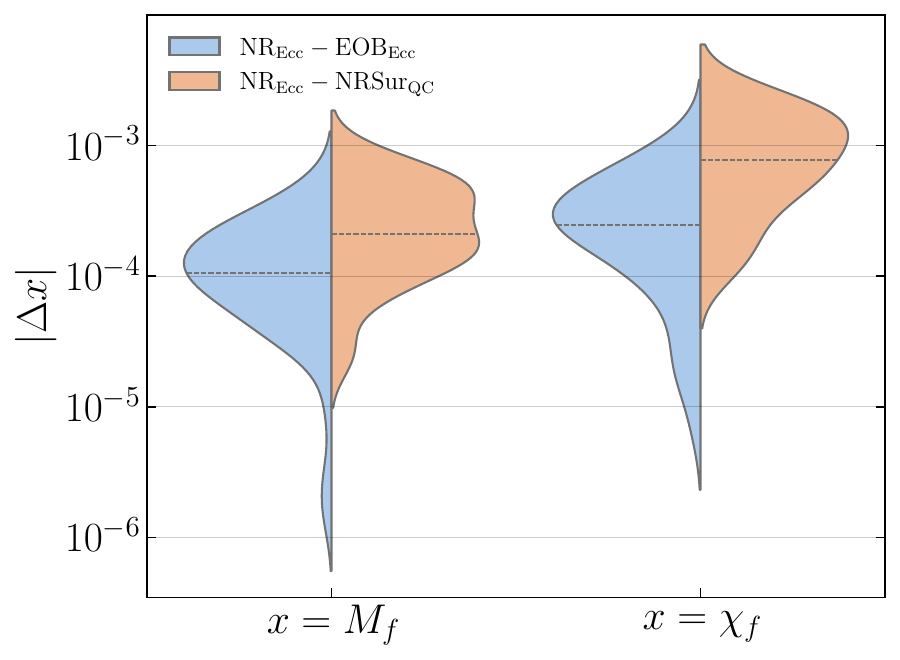}
    \caption{\textit{EOB estimates of the final mass and spin.} \textbf{Left:}
        Result of~(\ref{eq:final_mass_eob_ecc}) plotted with the actual NR data already shown in figure~\ref{plot: mean ano dependence}. \textbf{Right:} Absolute value of the difference between the NR and EOB predictions for the remnant mass and spin (in blue) across all NR simulations presented in this work, along with comparisons to the quasi-circular \texttt{NRSur7dq4Remnant} fit (in orange). Horizontal dashed lines indicate the median values.
    }
    \label{plot:eob_nr_comparisons}
\end{figure*}

The results for a subset of NR simulations with different values of mean anomaly $ \ell_{ \mathrm{gw}} $, but equal values of eccentricity $ e _{ \mathrm{gw}} \simeq 0.17 $ at a reference frequency $\langle\omega_{22}\rangle=0.069/M$, are presented in the left panel of figure~\ref{plot:eob_nr_comparisons}.
In this plot, we include the values of the final mass $ M_f$ extracted from the NR simulations, and the corresponding \texttt{SEOBNRv5EHM} estimates based on~\eqref{eq:final_mass_eob_ecc}.
For the EOB data points, the values of $ e _{\mathrm{gw}} $ and $ \ell_{\mathrm{gw}}$ are extracted with the \texttt{gw\_eccentricity} package at $\langle\omega_{22}\rangle$ from the optimum \texttt{SEOBNRv5EHM} waveforms.
We observe that the oscillatory dependence of $M_f$ around the quasi-circular value is also captured by the EOB estimates, with the amplitude and phasing being reasonably well reproduced.

To better assess the accuracy of the estimates (\ref{eq:final_mass_eob_ecc}) and (\ref{eq:final_spin_eob_ecc}), the right panel of figure~\ref{plot:eob_nr_comparisons} shows the absolute difference between the NR and EOB predictions for the remnant mass and spin (in blue) across all NR simulations presented in this work, along with comparisons to the quasi-circular \texttt{NRSur7dq4Remnant} fit (in orange). Horizontal dashed lines indicate the median values. For this comparison, the EOB estimates use the input parameters corresponding to the optimum waveforms.
The quasi-circular NR surrogate for the remnant properties does not capture eccentricity-induced variations as those shown in the left panel of figure~\ref{plot:eob_nr_comparisons}, whereas the EOB estimates reproduce such variations to a reasonable degree.  Therefore, the EOB-estimates reduce differences $|\Delta x|$ in the right panel of figure~\ref{plot:eob_nr_comparisons}  by a factor $\sim 2$ for the final mass and $\sim 3$ for the final spin.

We note that Eqs.~\eqref{eq:final_mass_eob_ecc} and \eqref{eq:final_spin_eob_ecc} can be evaluated during waveform generation with the \texttt{SEOBNRv5EHM} model at negligible computational cost, so that these corrections can be included without requiring additional parameter space fits.
It would also be straightforward to produce fits for these corrections using the same techniques applied to NR remnant fits, given the efficiency of generating large numbers of EOB waveforms.

\section{Conclusion}
\label{Sec:Conclusion}

In this work, we have presented a complete analysis of eccentric
simulations where, for the first time, both eccentric parameters are
accounted for. To do so, we propose a new waveform-based eccentricity
control method, which we implement in the Spectral Einstein Code
\texttt{SpEC}. This eccentricity control procedure enables a new,
precise control over both reference eccentricity and mean anomaly. We
utilize this procedure to generate several sequences of full NR simulations with constant reference eccentricity, but
uniform coverage of mean anomaly. We show that, for the parameter
space explored, merger-related quantities show an oscillatory
dependence on mean anomaly with mean very close to the corresponding quasi-circular
value, with the amplitude of these oscillations determined by the
eccentricity of the system (see, e.g.~figures~\ref{plot: mean ano dependence} and \ref{plot: ecc dependence}).

Our results highlight the importance in considering the entire 2-dimensional parameter space when studying eccentric systems. Several recent studies have found a seemingly oscillatory dependence of several quantities on eccentricity. By recreating a typical sequence of simulations used in such studies, we demonstrate that the reported oscillations arise from a lack of control over the reference mean anomaly of these systems.

While there are several subtleties in regards to defining reference eccentricity and mean anomaly for a binary, it is shown that so long as one defines these values sufficiently close to merger, the resulting dependence on both eccentricity and mean anomaly is quite simple. This result is promising in the context of waveform modelling, where the inclusion of such effects should be straightforward.

Finally, we show that some of the results in this work may be accessible by analytic models. By considering a previously used phenomenological estimate for the mass of the remnant BH, we show that the \texttt{SEOBNRv5EHM} model reproduces oscillations of a similar order of magnitude. This hints at a possible analytic explanation to some of our results, as was also recently explored in \cite{Wang:2024jro}.

There are many other quantities one can examine using our simulations. Continuing to focus on the merger portion of the evolution, the modification of the mass and spin of the remnant indicate that accounting for both the anomaly and eccentricity is necessary to characterise ringdowns from eccentric systems. Further, the variation of the peak of $A_{22}$ suggests that the excitation coefficients will likely have a similar dependence on both parameters. This is further supported by the notion that, in linear perturbation theory, the excitation coefficients depend on the exact perturbation imparted on the BH~\cite{Sun:1988tz}. As the mean anomaly picks out the ``merger geometry'' obtained, one should expect this will also choose the exact type of perturbation present in the remnant BH. We leave further exploration of these quantities to future works.

It is also interesting to consider the impact of eccentricity and mean anomaly on the recoil kick imparted on the remnant BH due to asymmetric emission of linear momentum~\cite{Radia:2021hjs}. Such a study, however, comes with some subtle difficulties; for systems of sufficient eccentricity, one obtains non-negligible kicks imparted on the binary system during \textit{inspiral} with each periastron passage. While these kicks are significantly smaller than that imparted on the remnant, their contributions make it more difficult to define the initial rest-frame which we relate the recoil to. As well as this, periastron precession leads to these kicks being imparted in varying directions, which can lead to both constructive and destructive contributions. We leave a more complete exploration of BH recoil to future work.

To account for mean anomaly in modern waveform models, one would have to explore the phenomenology of mean anomaly/eccentricity deviations across more of the parameter space. In particular, one would have to explore the impact of varying the mass-ratio and spins. Preliminary results indicate that the phenomenology of these deviations is similar in other parts of parameter space, indicating that fewer simulations may be required to parametrize these effects. We leave a more complete exploration of the non-precessing parameter space to future work.

While the new eccentricity control procedure presented is sufficient for our purposes, there are several avenues for improvement. Currently, we compare a finite-radius NR waveform to an \texttt{SEOBNRv5EHM} waveform associated with future null infinity. Yet more accurate control than what we achieved here may require to extrapolate the NR waveform to future null infinity~\cite{Boyle:2019kee}. One could also attempt to extend the eccentricity control procedure to generic spins, once analytic waveform models with both eccentricity and precessing spins are available.

\section{Acknowledgements}

The authors would like to acknowledge Katerina Chatziioannou, Nihar Gupte, Cheng Foo, Taylor Knapp, Benjamin Leather, Oliver Long, Philip Lynch, Maarten van de Meent, and Raj Patil for fruitful discussions. P.J.~Nee would like to thank many members of the ACR division for feedback regarding the phenomenology of the findings presented. Computations were performed on the Urania HPC system at the Max Planck Computing and Data Facility and the Expanse HPC system at the San Diego Supercomputer Center~\cite{10.1145/3437359.3465588}. A.~Ramos-Buades is supported by the Veni research programme which is (partly) financed by the Dutch Research Council (NWO) under the grant VI.Veni.222.396; acknowledges support from the Spanish Agencia Estatal de Investigación grant PID2022-138626NB-I00 funded by MICIU/AEI/10.13039/501100011033 and the ERDF/EU; is supported by the Spanish Ministerio de Ciencia, Innovación y Universidades (Beatriz Galindo, BG23/00056) and co-financed by UIB.
V.~Varma acknowledges support from NSF Grant No. PHY-2309301 and UMass Dartmouth's
Marine and Undersea Technology (MUST) Research Program funded by the Office of
Naval Research (ONR) under Grant No. N00014-23-1-2141.

\appendix
\section*{Appendix}
\label{sec:appendix}
Table~\ref{tab:uniform_runs} lists the SXS IDs of the simulations
analysed in section~\ref{sec:31}. In addition, we provide the initial
conditions $(D_0,\Omega_0,\dot{a}_0)$ used to perform these
simulations, which were obtained via the eccentricity control
procedure described in section~\ref{Sec:EOB-EccControl}.
The values $e^{\rm target}_0$ in
Table~\ref{tab:uniform_runs} refer to the initial eccentricities
reported by the EOB-model utilized in the eccentricity tuning
procedure.  We extract the actual eccentricities of each NR waveform
using \texttt{gw\_eccentricity} at reference frequency
$\langle\omega_{22}\rangle=0.03/M$, and plot their spread in figure~\ref{plot: ecc
    deviation}.  We find that the actual NR eccentricities are
constant within each sequence to fractional accuracy of a few $10^{-3}$.

\begin{figure}[b!]
    \centerline{\includegraphics[width=0.9\linewidth]{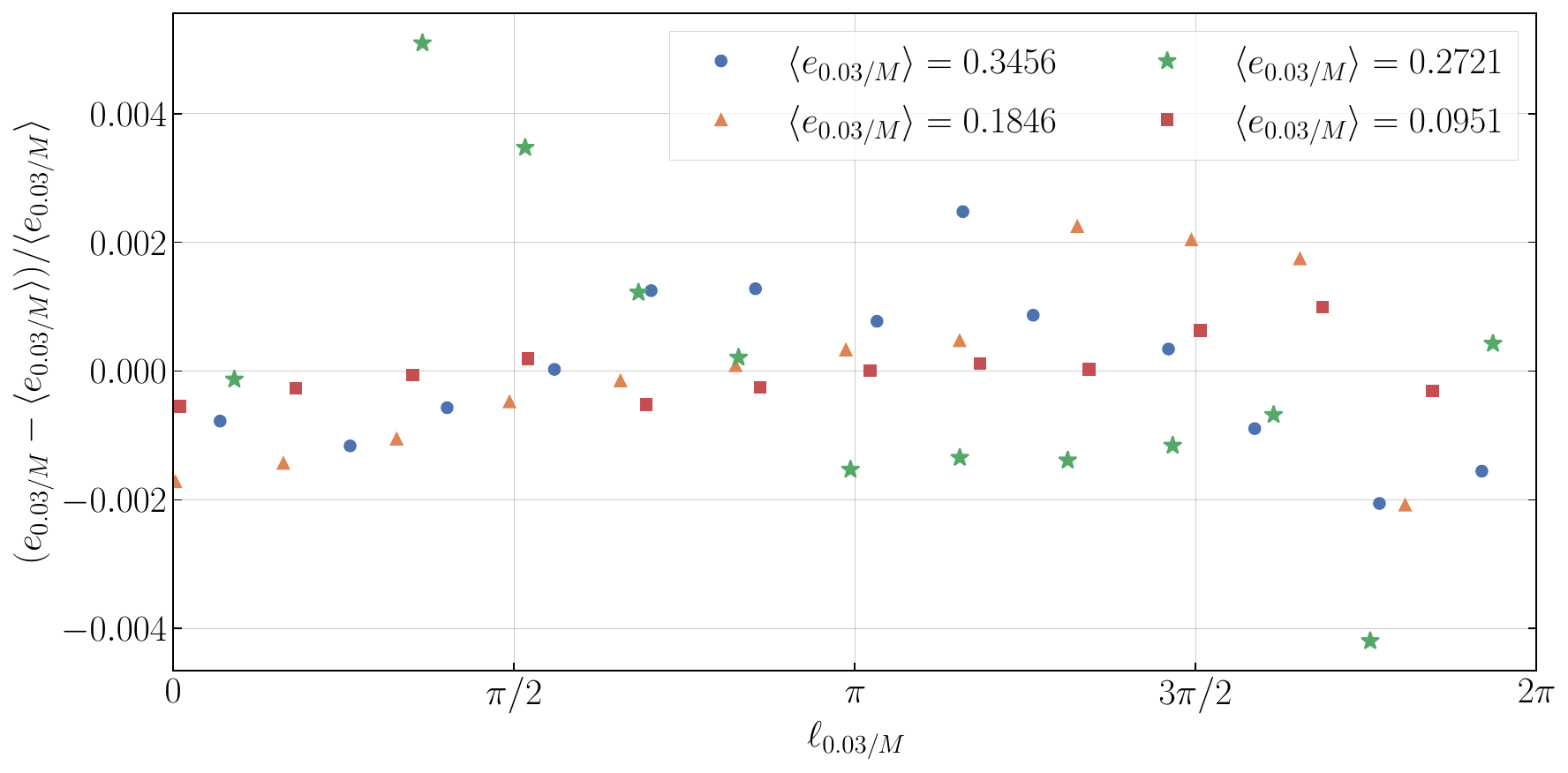}}
    \caption{\textit{Deviation of eccentricity within each constant-eccentricity sequence.} Each colour corresponds to one of the constant eccentricity sequences presented in figure~\ref{plot: A22_polar}. $\langle e_{0.03/M}\rangle$ is computed by taking the average of $e_{0.03/M}$ across each sequence.}

    \label{plot: ecc deviation}
\end{figure}

Table~\ref{tab:sequence} presents the initial data paramters for the simulations analysed in section~\ref{sec:32} (this analysis does not utilize eccentricity control).
\renewcommand{\arraystretch}{1.4}
\begin{center}
    \begin{longtable}{c|c|c|c|c|c}
        \hline\hline
        SXS ID       & $D_0/M$  & $M\Omega_0$            & $M\dot{a_0}$           & $e_0^{\rm target}$ & $T_{\rm merger}^{\rm target}/M$ \\\hline\hline
        \endfirsthead
        \hline\hline
        SXS ID       & $D_0/M$  & $M\Omega_0$            & $M\dot{a_0}$           & $e_0^{\rm target}$ & $T_{\rm merger}^{\rm target}/M$ \\\hline\hline
        \endhead
        \hline\hline
        \endfoot
        \endlastfoot
        SXS:BBH:4381 & $38.540$ & $3.032 \times 10^{-3}$ & $2.41 \times 10^{-6}$  & $0.5000$           & $11915$                         \\ \hline
        SXS:BBH:4293 & $38.527$ & $2.982 \times 10^{-3}$ & $2.07 \times 10^{-6}$  & $0.5000$           & $11884$                         \\ \hline
        SXS:BBH:4304 & $38.336$ & $3.009 \times 10^{-3}$ & $2.14 \times 10^{-6}$  & $0.4984$           & $11798$                         \\ \hline
        SXS:BBH:4303 & $38.207$ & $3.027 \times 10^{-3}$ & $2.19 \times 10^{-6}$  & $0.4974$           & $11719$                         \\ \hline
        SXS:BBH:4302 & $38.072$ & $3.046 \times 10^{-3}$ & $2.23 \times 10^{-6}$  & $0.4964$           & $11642$                         \\ \hline
        SXS:BBH:4301 & $37.896$ & $3.071 \times 10^{-3}$ & $2.30 \times 10^{-6}$  & $0.4948$           & $11557$                         \\ \hline
        SXS:BBH:4300 & $37.736$ & $3.094 \times 10^{-3}$ & $2.37 \times 10^{-6}$  & $0.4937$           & $11478$                         \\ \hline
        SXS:BBH:4299 & $37.590$ & $3.116 \times 10^{-3}$ & $2.44 \times 10^{-6}$  & $0.4925$           & $11398$                         \\ \hline
        SXS:BBH:4298 & $37.449$ & $3.136 \times 10^{-3}$ & $2.49 \times 10^{-6}$  & $0.4912$           & $11318$                         \\ \hline
        SXS:BBH:4297 & $37.289$ & $3.160 \times 10^{-3}$ & $2.56 \times 10^{-6}$  & $0.4899$           & $11238$                         \\ \hline
        SXS:BBH:4296 & $37.137$ & $3.183 \times 10^{-3}$ & $2.63 \times 10^{-6}$  & $0.4886$           & $11158$                         \\ \hline
        SXS:BBH:4295 & $37.030$ & $3.198 \times 10^{-3}$ & $2.66 \times 10^{-6}$  & $0.4877$           & $11084$                         \\ \hline
        SXS:BBH:4294 & $36.977$ & $3.205 \times 10^{-3}$ & $2.68 \times 10^{-6}$  & $0.4877$           & $11021$                         \\ \hline
        SXS:BBH:4375 & $31.321$ & $4.446 \times 10^{-3}$ & $1.95 \times 10^{-5}$  & $0.3748$           & $12018$                         \\ \hline
        SXS:BBH:4374 & $31.258$ & $4.461 \times 10^{-3}$ & $1.99 \times 10^{-5}$  & $0.3743$           & $11955$                         \\ \hline
        SXS:BBH:4372 & $31.073$ & $4.507 \times 10^{-3}$ & $2.14 \times 10^{-5}$  & $0.3715$           & $11822$                         \\ \hline
        SXS:BBH:4373 & $31.068$ & $4.511 \times 10^{-3}$ & $2.19 \times 10^{-5}$  & $0.3705$           & $11876$                         \\ \hline
        SXS:BBH:4371 & $30.990$ & $4.528 \times 10^{-3}$ & $2.22 \times 10^{-5}$  & $0.3703$           & $11757$                         \\ \hline
        SXS:BBH:4370 & $30.909$ & $4.548 \times 10^{-3}$ & $2.31 \times 10^{-5}$  & $0.3692$           & $11692$                         \\ \hline
        SXS:BBH:4369 & $30.832$ & $4.567 \times 10^{-3}$ & $2.39 \times 10^{-5}$  & $0.3680$           & $11627$                         \\ \hline
        SXS:BBH:4380 & $30.750$ & $4.588 \times 10^{-3}$ & $2.49 \times 10^{-5}$  & $0.3669$           & $11556$                         \\ \hline
        SXS:BBH:4379 & $30.706$ & $4.598 \times 10^{-3}$ & $2.51 \times 10^{-5}$  & $0.3667$           & $11491$                         \\ \hline
        SXS:BBH:4378 & $30.622$ & $4.617 \times 10^{-3}$ & $2.54 \times 10^{-5}$  & $0.3665$           & $11363$                         \\ \hline
        SXS:BBH:4377 & $30.582$ & $4.626 \times 10^{-3}$ & $2.54 \times 10^{-5}$  & $0.3664$           & $11301$                         \\ \hline
        SXS:BBH:4376 & $30.656$ & $4.610 \times 10^{-3}$ & $2.56 \times 10^{-5}$  & $0.3664$           & $11425$                         \\ \hline
        SXS:BBH:4321 & $25.721$ & $6.452 \times 10^{-3}$ & $-2.15 \times 10^{-5}$ & $0.2500$           & $11882$                         \\ \hline
        SXS:BBH:4332 & $25.677$ & $6.470 \times 10^{-3}$ & $-2.17 \times 10^{-5}$ & $0.2496$           & $11822$                         \\ \hline
        SXS:BBH:4331 & $25.733$ & $6.450 \times 10^{-3}$ & $-2.17 \times 10^{-5}$ & $0.2495$           & $11940$                         \\ \hline
        SXS:BBH:4330 & $25.629$ & $6.489 \times 10^{-3}$ & $-2.18 \times 10^{-5}$ & $0.2493$           & $11762$                         \\ \hline
        SXS:BBH:4329 & $25.562$ & $6.518 \times 10^{-3}$ & $-2.18 \times 10^{-5}$ & $0.2483$           & $11700$                         \\ \hline
        SXS:BBH:4328 & $25.512$ & $6.539 \times 10^{-3}$ & $-2.19 \times 10^{-5}$ & $0.2478$           & $11641$                         \\ \hline
        SXS:BBH:4327 & $25.462$ & $6.559 \times 10^{-3}$ & $-2.19 \times 10^{-5}$ & $0.2472$           & $11581$                         \\ \hline
        SXS:BBH:4326 & $25.412$ & $6.581 \times 10^{-3}$ & $-2.19 \times 10^{-5}$ & $0.2467$           & $11522$                         \\ \hline
        SXS:BBH:4325 & $25.362$ & $6.602 \times 10^{-3}$ & $-2.20 \times 10^{-5}$ & $0.2461$           & $11462$                         \\ \hline
        SXS:BBH:4324 & $25.309$ & $6.624 \times 10^{-3}$ & $-2.20 \times 10^{-5}$ & $0.2455$           & $11403$                         \\ \hline
        SXS:BBH:4323 & $25.258$ & $6.646 \times 10^{-3}$ & $-2.21 \times 10^{-5}$ & $0.2450$           & $11344$                         \\ \hline
        SXS:BBH:4322 & $25.210$ & $6.667 \times 10^{-3}$ & $-2.21 \times 10^{-5}$ & $0.2445$           & $11285$                         \\ \hline
        SXS:BBH:4361 & $22.012$ & $8.639 \times 10^{-3}$ & $-1.79 \times 10^{-5}$ & $0.1248$           & $11936$                         \\ \hline
        SXS:BBH:4360 & $21.979$ & $8.660 \times 10^{-3}$ & $-1.80 \times 10^{-5}$ & $0.1245$           & $11881$                         \\ \hline
        SXS:BBH:4359 & $21.948$ & $8.678 \times 10^{-3}$ & $-1.81 \times 10^{-5}$ & $0.1243$           & $11826$                         \\ \hline
        SXS:BBH:4358 & $21.916$ & $8.698 \times 10^{-3}$ & $-1.81 \times 10^{-5}$ & $0.1241$           & $11771$                         \\ \hline
        SXS:BBH:4357 & $21.884$ & $8.717 \times 10^{-3}$ & $-1.82 \times 10^{-5}$ & $0.1239$           & $11716$                         \\ \hline
        SXS:BBH:4368 & $21.851$ & $8.738 \times 10^{-3}$ & $-1.83 \times 10^{-5}$ & $0.1236$           & $11658$                         \\ \hline
        SXS:BBH:4367 & $21.820$ & $8.756 \times 10^{-3}$ & $-1.84 \times 10^{-5}$ & $0.1234$           & $11601$                         \\ \hline
        SXS:BBH:4366 & $21.787$ & $8.777 \times 10^{-3}$ & $-1.85 \times 10^{-5}$ & $0.1232$           & $11543$                         \\ \hline
        SXS:BBH:4365 & $21.754$ & $8.797 \times 10^{-3}$ & $-1.86 \times 10^{-5}$ & $0.1229$           & $11486$                         \\ \hline
        SXS:BBH:4364 & $21.721$ & $8.818 \times 10^{-3}$ & $-1.87 \times 10^{-5}$ & $0.1227$           & $11429$                         \\ \hline
        SXS:BBH:4363 & $21.690$ & $8.837 \times 10^{-3}$ & $-1.87 \times 10^{-5}$ & $0.1225$           & $11374$                         \\ \hline
        SXS:BBH:4362 & $21.662$ & $8.853 \times 10^{-3}$ & $-1.88 \times 10^{-5}$ & $0.1225$           & $11319$                         \\ \hline\hline
        \caption{Simulations analysed in section~\ref{sec:31}. The columns show the SXS ID, as well as the initial conditions $(D_0,\Omega_0,\dot{a}_0)$ obtained from the procedure presented in section~\ref{Sec:EOB-EccControl}.} \label{tab:uniform_runs}
    \end{longtable}
\end{center}

\begin{center}
    \begin{longtable}{c|c|c|c}
        \hline\hline
        SXS ID       & $D_0/M$  & $M\Omega_0$            & $M\dot{a_0}$          \\\hline\hline
        \endfirsthead
        \hline\hline
        SXS ID       & $D_0/M$  & $M\Omega_0$            & $M\dot{a_0}$          \\\hline\hline
        \endhead
        \hline\hline
        \endfoot
        \endlastfoot
        SXS:BBH:4392 & $24.102$ & $7.203 \times 10^{-3}$ & $1.02 \times 10^{-5}$ \\ \hline
        SXS:BBH:4408 & $24.102$ & $7.181 \times 10^{-3}$ & $1.02 \times 10^{-5}$ \\ \hline
        SXS:BBH:4398 & $24.102$ & $7.158 \times 10^{-3}$ & $1.02 \times 10^{-5}$ \\ \hline
        SXS:BBH:4409 & $24.102$ & $7.136 \times 10^{-3}$ & $1.02 \times 10^{-5}$ \\ \hline
        SXS:BBH:4399 & $24.102$ & $7.114 \times 10^{-3}$ & $1.02 \times 10^{-5}$ \\ \hline
        SXS:BBH:4410 & $24.102$ & $7.092 \times 10^{-3}$ & $1.02 \times 10^{-5}$ \\ \hline
        SXS:BBH:4393 & $24.102$ & $7.069 \times 10^{-3}$ & $1.02 \times 10^{-5}$ \\ \hline
        SXS:BBH:4411 & $24.102$ & $7.047 \times 10^{-3}$ & $1.02 \times 10^{-5}$ \\ \hline
        SXS:BBH:4400 & $24.102$ & $7.025 \times 10^{-3}$ & $1.02 \times 10^{-5}$ \\ \hline
        SXS:BBH:4412 & $24.102$ & $7.002 \times 10^{-3}$ & $1.02 \times 10^{-5}$ \\ \hline
        SXS:BBH:4401 & $24.102$ & $6.980 \times 10^{-3}$ & $1.02 \times 10^{-5}$ \\ \hline
        SXS:BBH:4413 & $24.102$ & $6.958 \times 10^{-3}$ & $1.02 \times 10^{-5}$ \\ \hline
        SXS:BBH:4394 & $24.102$ & $6.936 \times 10^{-3}$ & $1.02 \times 10^{-5}$ \\ \hline
        SXS:BBH:4414 & $24.102$ & $6.913 \times 10^{-3}$ & $1.02 \times 10^{-5}$ \\ \hline
        SXS:BBH:4402 & $24.102$ & $6.891 \times 10^{-3}$ & $1.02 \times 10^{-5}$ \\ \hline
        SXS:BBH:4415 & $24.102$ & $6.869 \times 10^{-3}$ & $1.02 \times 10^{-5}$ \\ \hline
        SXS:BBH:4403 & $24.102$ & $6.847 \times 10^{-3}$ & $1.02 \times 10^{-5}$ \\ \hline
        SXS:BBH:4416 & $24.102$ & $6.824 \times 10^{-3}$ & $1.02 \times 10^{-5}$ \\ \hline
        SXS:BBH:4395 & $24.102$ & $6.802 \times 10^{-3}$ & $1.02 \times 10^{-5}$ \\ \hline
        SXS:BBH:4417 & $24.102$ & $6.780 \times 10^{-3}$ & $1.02 \times 10^{-5}$ \\ \hline
        SXS:BBH:4404 & $24.102$ & $6.757 \times 10^{-3}$ & $1.02 \times 10^{-5}$ \\ \hline
        SXS:BBH:4418 & $24.102$ & $6.735 \times 10^{-3}$ & $1.02 \times 10^{-5}$ \\ \hline
        SXS:BBH:4405 & $24.102$ & $6.713 \times 10^{-3}$ & $1.02 \times 10^{-5}$ \\ \hline
        SXS:BBH:4419 & $24.102$ & $6.691 \times 10^{-3}$ & $1.02 \times 10^{-5}$ \\ \hline
        SXS:BBH:4396 & $24.102$ & $6.668 \times 10^{-3}$ & $1.02 \times 10^{-5}$ \\ \hline
        SXS:BBH:4420 & $24.102$ & $6.646 \times 10^{-3}$ & $1.02 \times 10^{-5}$ \\ \hline
        SXS:BBH:4406 & $24.102$ & $6.624 \times 10^{-3}$ & $1.02 \times 10^{-5}$ \\ \hline
        SXS:BBH:4421 & $24.102$ & $6.602 \times 10^{-3}$ & $1.02 \times 10^{-5}$ \\ \hline
        SXS:BBH:4407 & $24.102$ & $6.579 \times 10^{-3}$ & $1.02 \times 10^{-5}$ \\ \hline
        SXS:BBH:4422 & $24.102$ & $6.557 \times 10^{-3}$ & $1.02 \times 10^{-5}$ \\ \hline
        SXS:BBH:4397 & $24.102$ & $6.535 \times 10^{-3}$ & $1.02 \times 10^{-5}$ \\ \hline\hline
        \caption{Simulations analysed in section~\ref{sec:32}. The columns show the SXS ID, as well as the initial conditions $(D_0,\Omega_0,\dot{a}_0)$ used to perform these simulations.} \label{tab:sequence}
    \end{longtable}
\end{center}

\setcounter{section}{1}
\section*{References}
\bibliographystyle{iopart_num}
\bibliography{merged}

\end{document}